\documentclass[preprint, raferee]{aastex}
\usepackage{epsfig,amsmath,amsfonts,amssymb,graphicx,subfigure,enumitem,color}
\usepackage{amssymb}
\usepackage{amsmath}
\usepackage[varg]{txfonts}
\usepackage{natbib}
\usepackage{url}             
\usepackage{graphicx}
\usepackage{xcolor}
\usepackage[percent]{overpic}
\usepackage{mathptmx}
\usepackage{anyfontsize}
\usepackage{t1enc}
\usepackage{url}
\usepackage{float}
\usepackage[T1]{fontenc}
\usepackage[utf8]{inputenc}
\usepackage[spanish]{babel}

\newcommand{\km}{{\ \mathrm {km}} }
\newcommand{\K}{{\ \mathrm K} }
\newcommand{\s}{{\ \mathrm s} }
\def\arcsec{\hbox{$^{\prime\prime}$}}

\graphicspath{{./}{figures/}}
\usepackage{float}
\newcommand{\beq}{\begin{equation}}
\newcommand{\eeq}{\end{equation}}
\newcommand{\beqa}{\begin{eqnarray}}
\newcommand{\eeqa}{\end{eqnarray}}
\newcommand{\beqann}{\begin{eqnarray*}}
\newcommand{\eeqann}{\end{eqnarray*}}
\bibpunct{(}{)}{;}{a}{}{,} 

\usepackage{textcomp}
\shorttitle{Plasma Flows}
\shortauthors{Rao et al.}
\begin{document}
\title{Plasma flows in the cool loop systems}

\author{Yamini K. Rao$^{1}$, Abhishek K. Srivastava$^{1}$, Pradeep Kayshap$^{2}$, Klaus Wilhelm$^{3}$, Bhola N. Dwivedi$^{1}$}
\affil{$^{1}$Department of Physics, Indian Institute of Technology (BHU), Varanasi-221005, India.}
\affil{$^{2}$Institute of Physics, University of South Bohemia, Brani\v sovsk\'a 1760, CZ -- 370 05 \v{C}esk\'e Bud\v{e}jovice, Czech Republic.}
\affil{$^{3}$Max-Planck-Institut f\"ur Sonnensytemforschung, G\"ottingen, Justus-von-Liebig-Weg 3, 37077, Germany}

\bigskip
\begin{abstract}
We study the dynamics of low-lying cool loop systems 
for three datasets as observed by the Interface Region Imaging Spectrograph (IRIS). 
Radiances, Doppler shifts and line widths are investigated in 
and around observed cool loop systems using various 
spectral lines formed between the photosphere and transition region (TR). 
Footpoints of the loop threads are either dominated by
blueshifts or redshifts. The co-spatial variation of velocity above the blue-shifted footpoints of various loop threads shows a transition from very small upflow velocities ranging from (-1 {\rm to} +1) $\km\s^{-1}$ in the Mg\,{\sc ii} k line
(2796.20~\AA; formation temperature: $\log (T/\K)$ = 4.0) to the high upflow velocities from (-10 {\rm to} -20) $\km\s^{-1}$ in Si\,{\sc iv}. Thus, the transition of the plasma flows
from red-shift (downflows) to the blue-shift (upflows) is observed above the footpoints of these loop systems in the spectral line
C\,{\sc ii} (1334.53~\AA; $\log (T/\K)$ = 4.3) lying between 
Mg\,{\sc ii} k and Si\,{\sc iv} (1402.77~\AA; $\log (T / K)$ = 4.8).
This flow inversion is consistently observed in all three sets of the observational data.
The other footpoint of loop system always remains red-shifted indicating downflowing plasma.
The multi-spectral line analysis in the present paper provides a 
detailed scenario of the plasma flows inversions in cool loop systems leading to the mass transport 
and their formation. The impulsive energy release due to small-scale reconnection above loop footpoint seems to be the most likely cause for sudden initiation of the plasma flows evident at TR temperatures.
\end{abstract}


\section{Introduction} 
\label{sec:intro}
Physics of the transition region (TR) of the Sun plays an important role in understanding the various dynamical plasma processes. It acts as an interface layer between the chromosphere and the lower corona. The closed magnetic structures are embedded in plasma
in different regions of the Sun, e.g., quiet-Sun (QS) and active regions (ARs). The loops are anchored in the solar photosphere while their upper segments
may lie either in the chromosphere and TR (low-lying loops), or in the corona.
The magnetic loops are classified on the basis of their formation
temperatures and emissions. There are hot loops (Dowdy 1993),
warm loops (Lenz et al. 1999)
, and cool loops (Foukal 1976)
having temperatures ranging
from $>$2 MK, (1~to~2)~MK, and (0.1~to~1.0)~MK, respectively.
This classification also depends on the spectral regime, i.e, cool loops are visible at
UV wavelengths,
warm loops in extreme ultraviolet (EUV) wavelengths, and hot loops emit
ultraviolet (UV), EUV, and X-ray wavelengths.
But even cooler loops, possessing temperatures lower
than 0.1 MK, 
also contribute significantly to the EUV emissions
(Feldman 1998; Sasso et al. 2012 and references cited therein). There are different scenarios proposed for
the heating of such loops, e.g., steady heating, impulsive heating,
etc. (Klimchuk 2006).

The loop structures in the solar atmosphere are the manifestations of
magnetic structures along which the plasma is confined.
In order to understand the heating of loop plasma, 
thermal diagnostics have to be conducted (Landi 2007).
Though the
corona sometimes flares in active regions,
these loops are  mostly considered to remain in a steady state (Rosner et al. 1978).
This shows that the heating mechanism has to be steady enough to
bring loops in an equilibrium condition. Also, the emission in coronal loops
varies significantly, which is found to be more sensitive
to density and less to temperature (Benz 2008).
Thus, it shows that the diagnostics of flows in coronal loops is not that simple
because brightness variations can even be produced by thermal fronts and
waves (Reale 2014).
Doppler shift measurements for different spectral lines
along the line-of-sight direction thus provides some
evidence of plasma motions. 
This supports mainly two types of bulk mass motions:
siphon flow often observed in the cool loops due to
pressure difference at different foot points or loop filling and draining, due to momentary heating and cooling, respectively.
Significant downflows are driven by cooling for which
two mechanisms have been proposed, viz: cool downfalling blobs of plasma
or slow magnetoacoustic waves (Bradshaw \& Cargill 2005; Cargill \& Bradshaw 2013).

\begin{figure*}
\begin{center}
\mbox{
\hspace{-1.5cm}
\includegraphics[trim=5.0cm 0.5cm 4.5cm 0.5cm,scale=0.85]{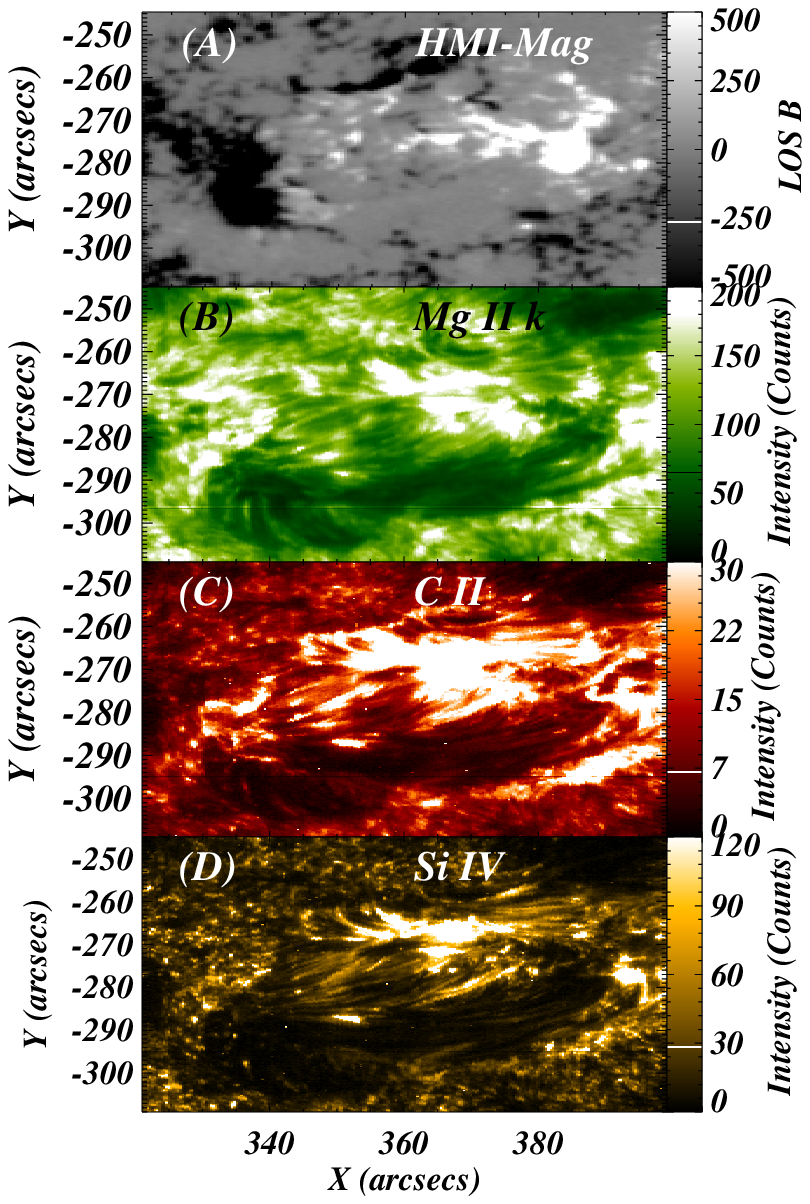}
\includegraphics[trim=5.0cm 0.5cm 4.5cm 0.5cm,scale=0.85]{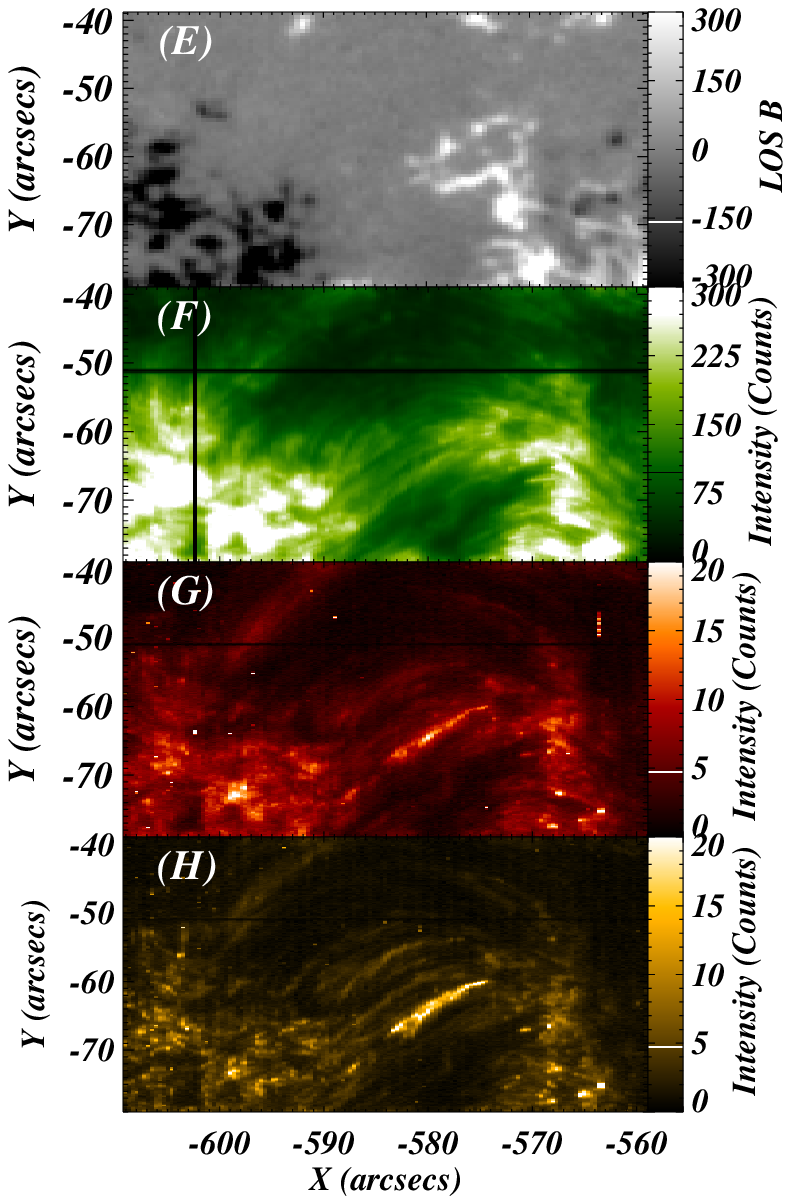}
\includegraphics[trim=5.0cm 0.5cm 4.5cm 0.5cm,scale=0.85]{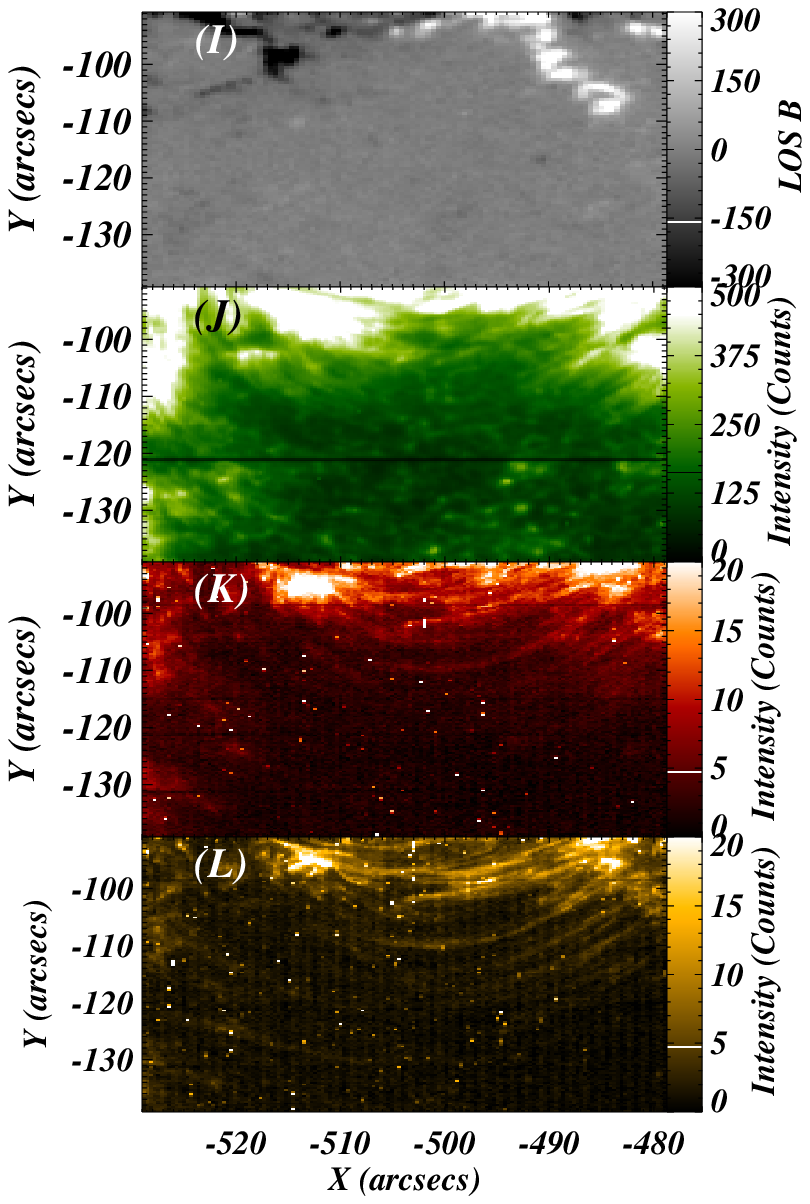}
}
\caption{The region of interest displaying the cool loop system in different
spectral lines: Mg\,{\sc ii}\,k (2796.20~\AA),
C\,{\sc ii} (1334.53~\AA), and
Si\,{\sc iv} (1402.77~\AA) along with the underlying magnetic
polarities indicated by HMI LOS magnetogram are shown for three different datasets. The left column corresponds to Dataset 1 observed on 27th December 2013 targeting AR 11934. The middle column indicates Dataset 2 observed on 10th December 2015 targeting AR 12465. The right column shows the Dataset 3 observed on 29th March 2017 targeting AR 12645.}
\label{fig_mosaic1}
\end{center}
\end{figure*}

Recently, Huang et al. (2015) have provided observational evidence
of cool TR loops using Si\,{\sc iv} (1402.77~\AA)
observations of the
Interface Region Imaging Spectrograph (IRIS).
However, very few studies have been conducted on cool loops due
to instrumental limitations. Earlier, cool loops have been observed using
spectral data from other instruments like the Solar Ultraviolet Measurements of
Emitted Radiation (SUMER) spectrograph on the Solar and Heliospheric Observatory
(SOHO) and the EUV Imaging Spectrometer (EIS) on the Hinode.
Doyle et al. (2006) have reported a redshift
of $\approx 20\km\s^{-1}$
in the N\,{\sc v} (1238.82~\AA) line at the footpoints of cool loops using
SUMER observations. Chae et al. (2000) have presented the rotational
motions along the axes of the loops having velocities of
$\approx 50\km\s^{-1}$
using H\,{\sc i} Ly$\beta$ (1025~\AA), O\,{\sc vi} (1032~\AA, 1038~\AA),
and C\,{\sc ii} (1037~\AA)
lines from SUMER in an AR.
The inversion of flows from redshift to blueshift has been observed in
different regions of the Sun (Warren et al. 1997, Brekke et al.
1997, Peter \& Judge 1999, Chae et al. 1998, Teriaca et al. 1999, Dadashi et al. 2011, Kayshap et al. 2015,
and references cited therein).
However, flows in such cool loop systems at the chromosphere/TR
interface have not been investigated in a deatiled manner.

The study of cool loops that are generally invisible at coronal temperatures,
and their plasma dynamics, are important candidates to understand correctly
the energy and mass transport processes in the solar atmosphere.
The heights at which such flows start to propagate upward have not yet been
properly understood. In order to understand the formations of cool loops,
their energetics, mass supply, the estimation of the flow and structure in such loops are required.
Inference of plasma flows in cool loop systems,
is essential to estimating flows above their footpoints at different heights
using different spectral lines having different
formation temperatures.
\begin{figure*}
\mbox{
\hspace{-0.5cm}
\includegraphics[trim=2.5cm 0.5cm 0.2cm 0.2cm,scale=0.6]{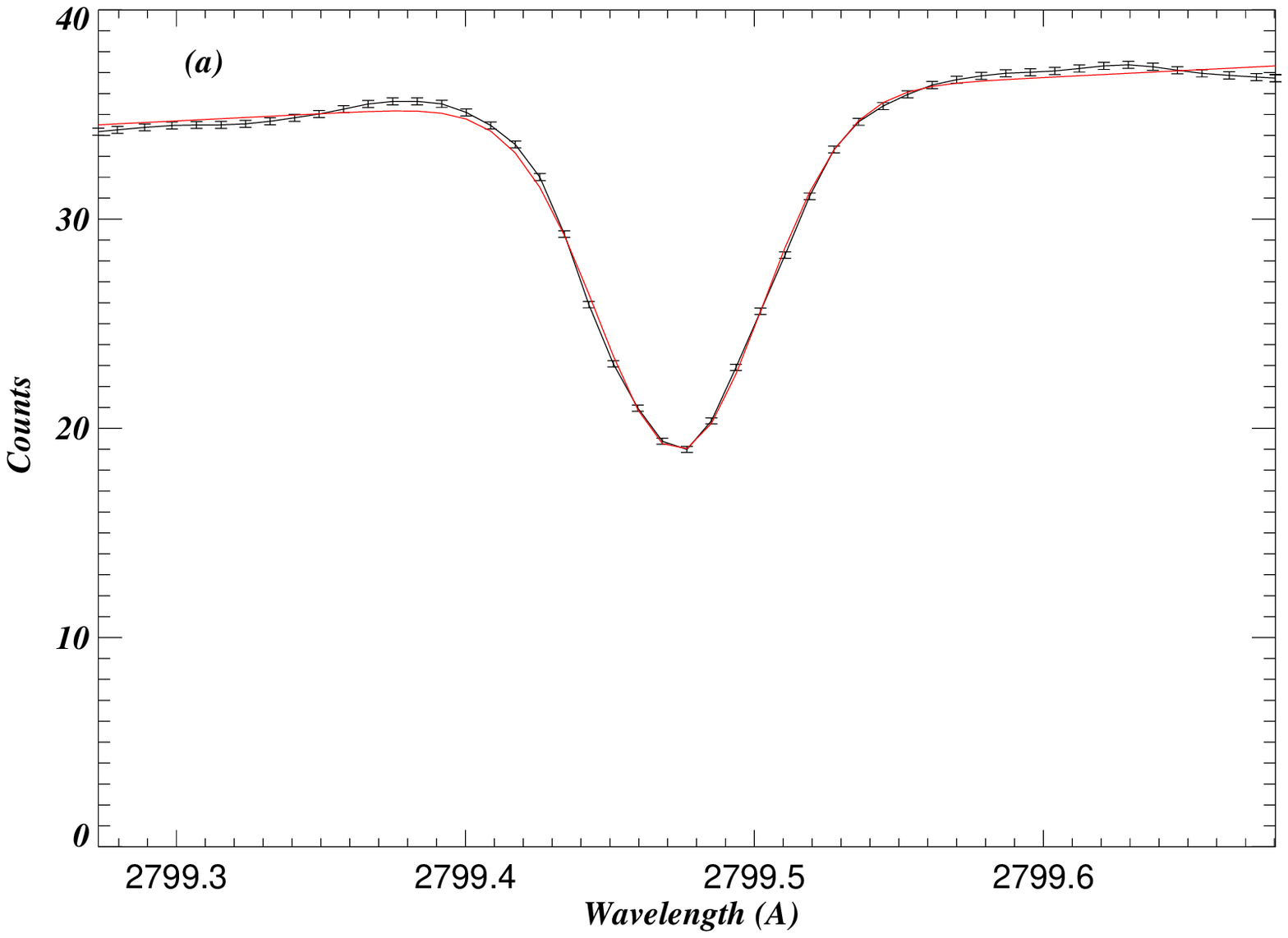}
\includegraphics[trim=1.0cm 0.5cm 0.2cm 0.2cm,scale=0.6]{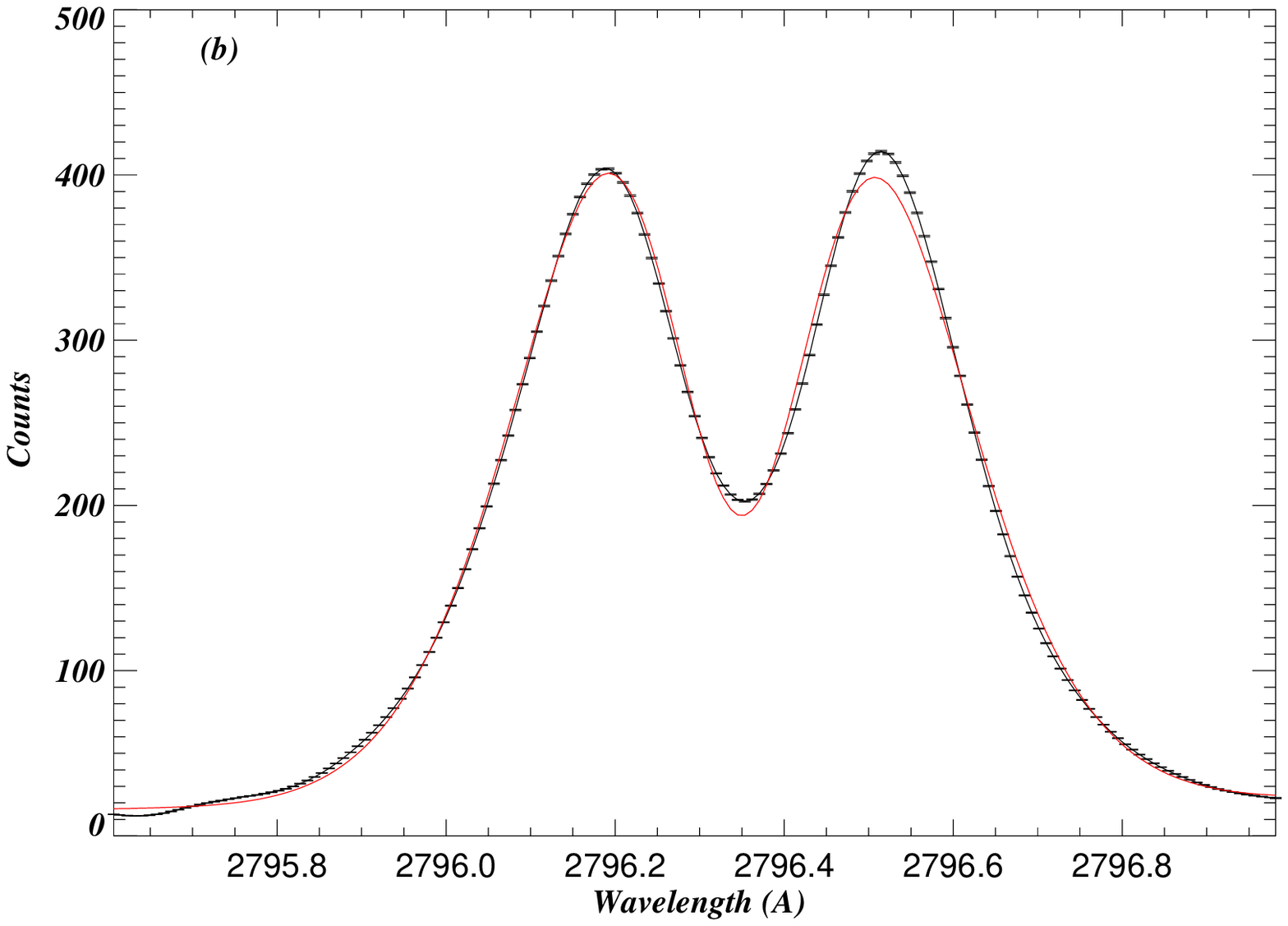}}
\mbox{
\hspace{-0.5cm}
\includegraphics[trim=2.5cm 0.5cm 0.2cm 0.2cm,scale=0.6]{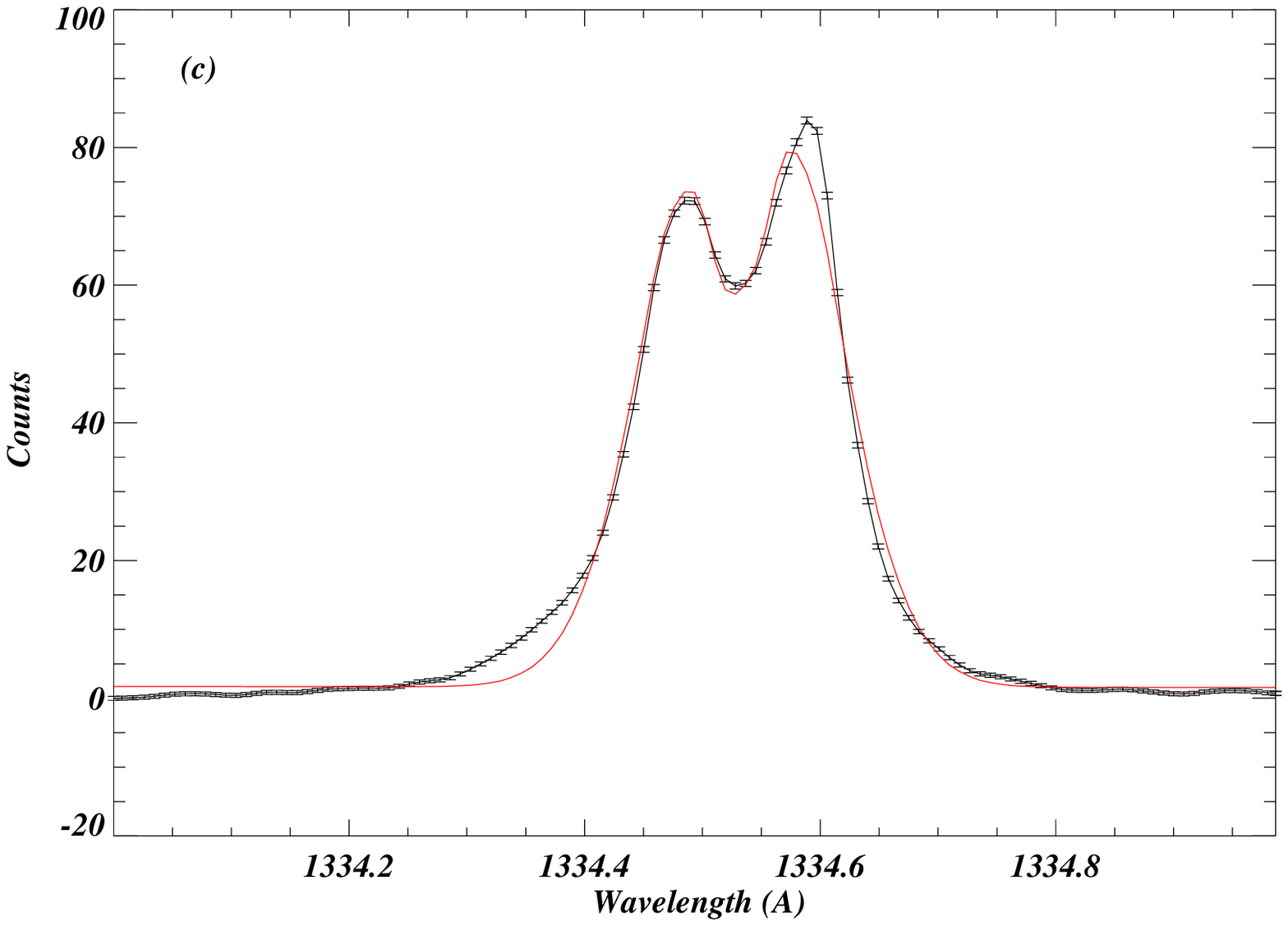}
\includegraphics[trim=1.0cm 0.5cm 0.2cm 0.2cm,scale=0.6]{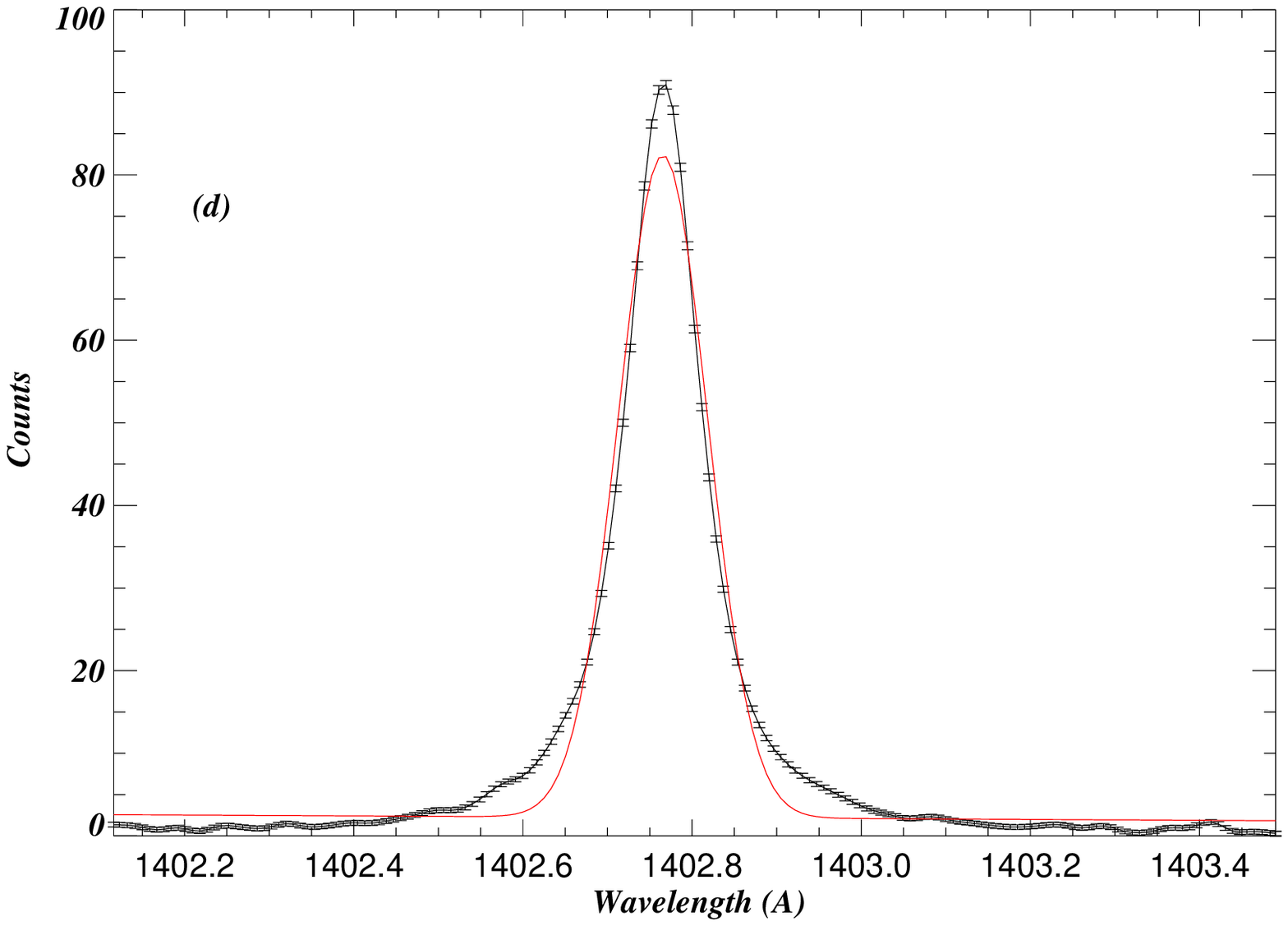}
}
\caption{Spectral fitting of (a) Ni\,{\sc I} (2799.47~\AA), (b) Mg\,{\sc II}\,k (2796.2~\AA), (c) C II (1354.53~\AA), and (d) Si\,{\sc IV} (1402.77~\AA) lines for averaged profile over the box labelled as B5 shown in the Si\,{\sc IV} left panel of Fig.~\ref{fig_vel1} corresponding to Dataset 1.}
\label{fig_spec1}
\end{figure*}

For our investigation in this work, we analyze similar regions using
three different datasets comprising of cool loop systems
using the spectral lines
Mg\,{\sc ii}\,k (2796.20~\AA) and Ni\,{\sc i} (2799.47~\AA) in the
near ultraviolet (NUV) domain
as well as in far ultraviolet (FUV) lines
C\,{\sc ii} (1334.53~\AA) and Si\,{\sc iv} (1402.77~\AA) observed by IRIS (De Pontieu et al. 2014).
We study the Doppler velocity pattern above the footpoints of
the cool loop systems as evident in three different spectral
observational datasets.
The co-spatial variation of Doppler velocities above the footpoints
of the cool loop systems provides the information about regions where these plasma flows have been triggered.
The other characteristic parameters such as radiance and FWHM have also been 
estimated to provide more insight of the response of the plasma flows in the cool loop systems. 
Consistency of the results is
described in three different epochs of the spectral observations
of cool loop systems in different parts of the Sun on different dates.
In Section~\ref{sec:obs_data}, we describe the
observational data and their analyses presenting 
details of all the three datasets which have been used.
Section~\ref{sec:results} describes the
results and their interpretation which have been classified in three different subsections of three different datasets.
In each subsection, identification of loops in various
spectral lines along with the magnetic polarities at their
footpoints\,--\,measured
with the Helioseismic and Magnetic Imager (HMI) on the
Solar Dynamics Observatory (SDO)\,--\,have been discussed. The corresponding
parametric maps are shown.
The variation of Doppler velocity at the footpoints of the
cool loop systems with different spectral lines has also been examined.
In the last section, discussions and
conclusions have been outlined.

\begin{figure*}
\mbox{
\includegraphics[trim=5.5cm 1.0cm 4.5cm 0.1cm,scale=0.85]{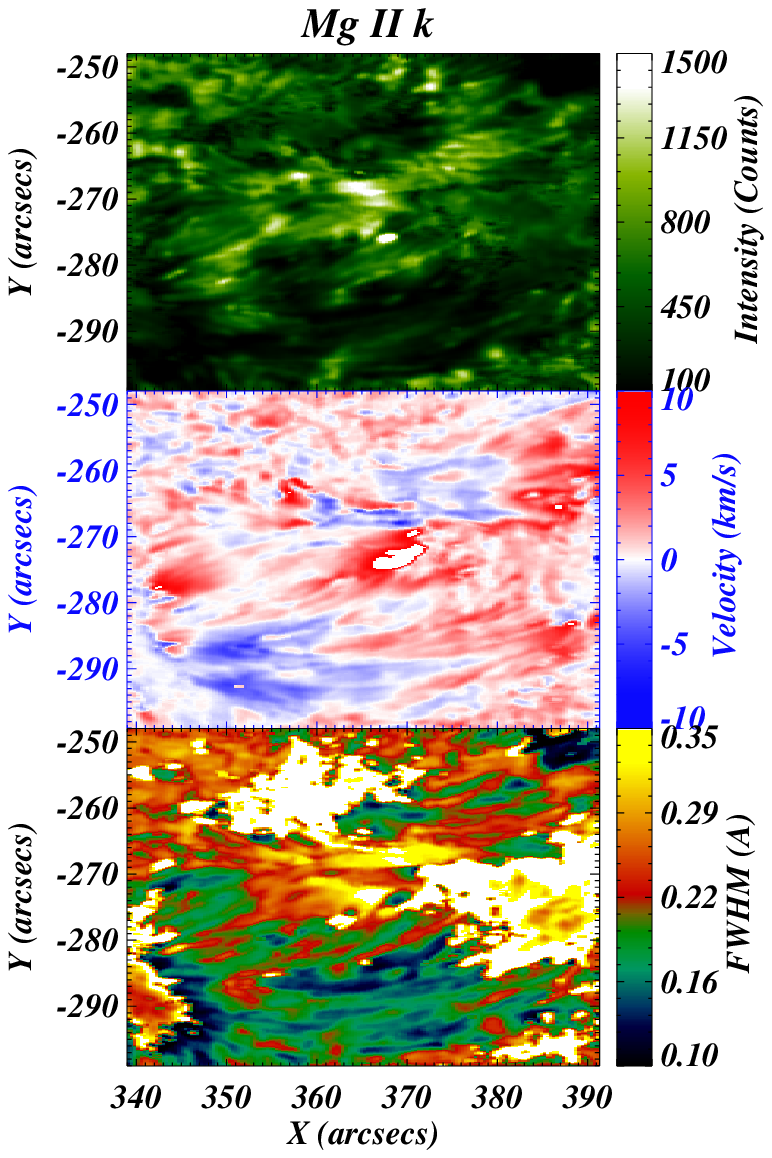}
\includegraphics[trim=5.5cm 1.0cm 4.5cm 0.1cm,scale=0.85]{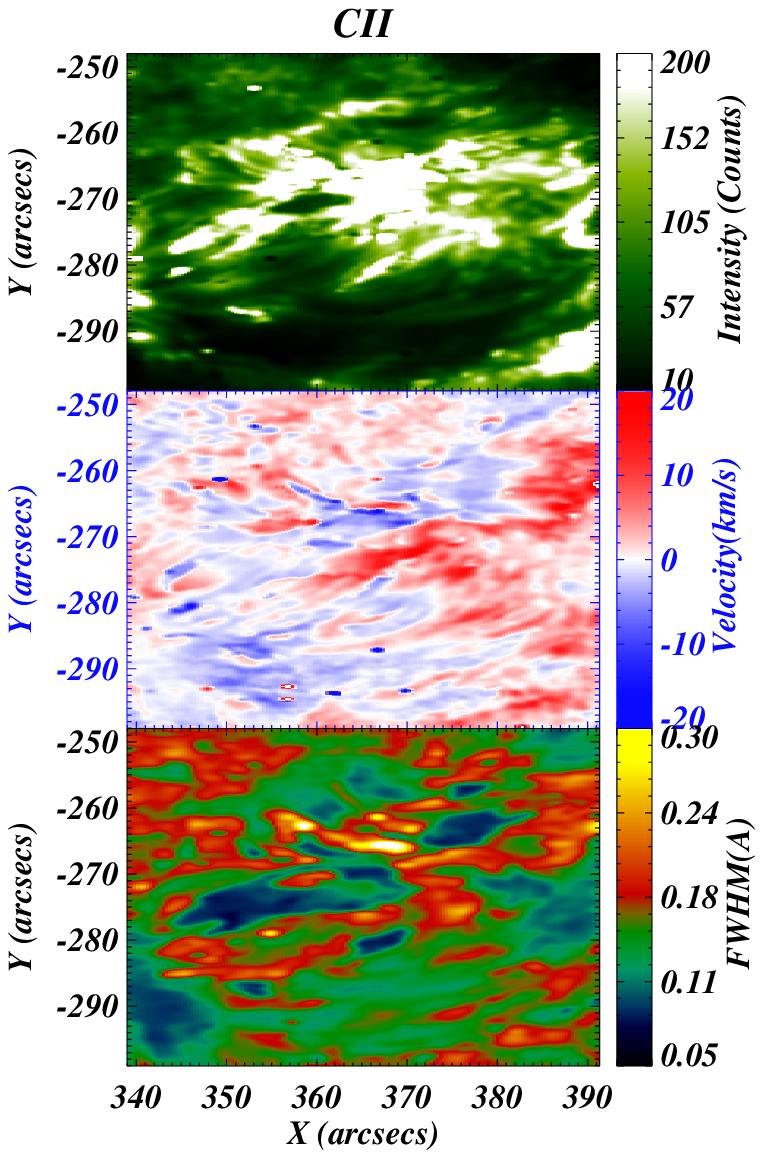}
\includegraphics[trim=5.5cm 1.0cm 4.5cm 0.1cm,scale=0.85]{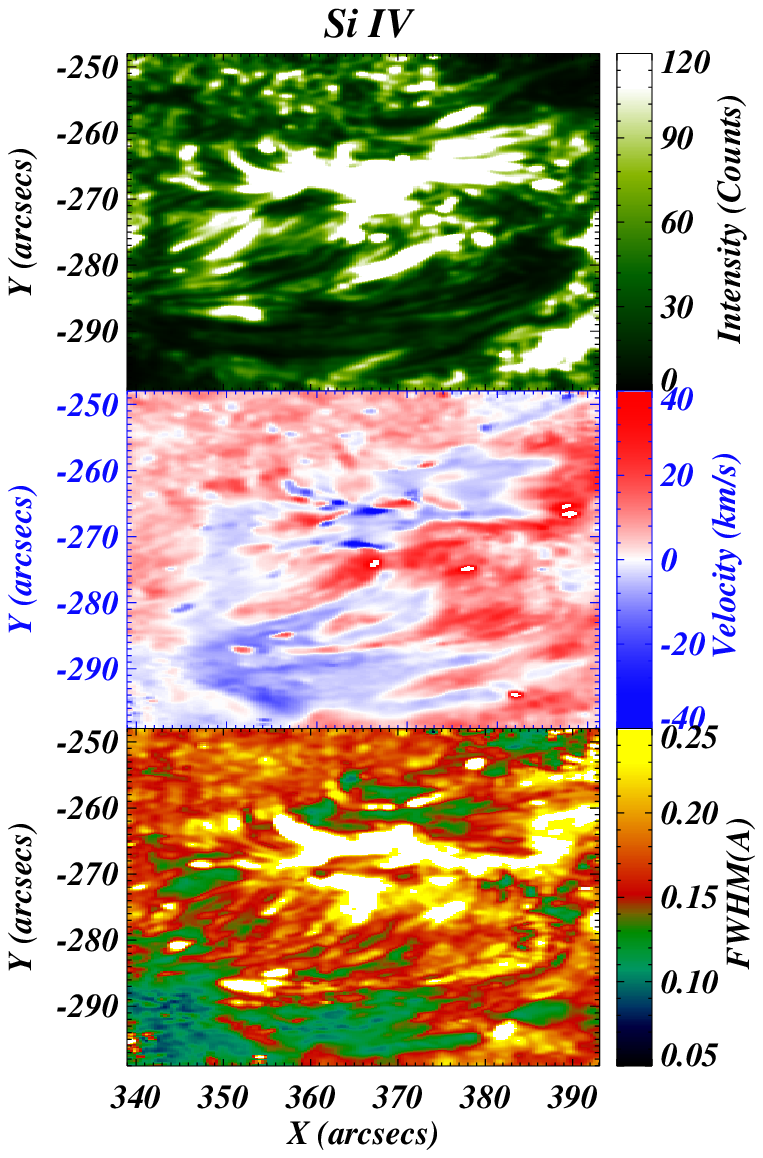}}
\caption{Intensity, Doppler velocity, and Full width at half maximum (FWHM) maps
of Mg\,{\sc II}\,k (2796.2~\AA) ,
C\,{\sc II} (1334.53~\AA), and
Si\,{\sc IV} (1402.77~\AA)
lines are shown for the Dataset 1
in the left, middle, and right columns respectively.}
\label{fig_data1}
\end{figure*}

\section{Observational data and their analyses} \label{sec:obs_data}
Spectra from IRIS provide data in the FUV band (1331.7~\AA~to 1358.4~\AA~and 1389.0~\AA~to 1407.0~\AA) and in the NUV (2782.7~\AA~to 2835.1~\AA) domain having a large number of spectral lines covering the photosphere, chromosphere, TR, and inner corona. Level~2 data are used for this study, which are calibrated for the dark current removal, flat fielding. We have utilized Si\,{\sc iv} (1402.77~\AA), Mg\,{\sc ii}\,k (2796.20~\AA), C\,{\sc ii} (1334.53~\AA) and Ni\,{\sc i} (2799.47~\AA) spectral lines. 
We study the cool loop system for three different datasets using these spectral lines. 
All the datasets used for our analysis have been chosen in such a way that the loop systems should have low connectivity with magneto-plasma threads visible in SJI 1400 \AA~. The loop systems connected with the outer periphery of an active region or its surroundings should not contain any flare disruption. Also, the loop systems should not contain any jet activity or disruption of field lines during the whole scan  of the raster maintaining the quiet nature of the loops.

Dataset 1: The dataset used for our study was observed by IRIS during the time
period from 22:38:08 to 23:11:59 UTC on 27th December 2013 targeting AR\,11934.
IRIS observed raster of  $141\arcsec$~in the $x$-direction and $174\arcsec$~in the
$y$-direction centred at ($X_{\rm cen},Y_{\rm cen}) = (341\arcsec,-267\arcsec)$.
The raster scan has a temporal cadence of 5.1$\s$. The whole scan of the raster has 400 pixels in the $x$-direction with the pixel size 0.35$\arcsec$~and 1096 pixels in the $y$-direction with the pixel size 0.16$\arcsec$. The region of interest
ranges from $321\arcsec~{\rm to}~399\arcsec$~in $x$-direction and $-309\arcsec~{\rm to}~-245\arcsec$~in $y$-direction (left-column (panel A,B,C andD); Fig.~\ref{fig_mosaic1}). \\
Dataset 2: The dataset used for our study was observed by IRIS during the time
period from 02:02:48 to 02:53:46 UTC on 10th December 2015 targeting AR\,12465.
IRIS observed raster of  $112\arcsec$~in the $x$-direction and $119\arcsec$~in the $y$-direction centred at ($X_{\rm cen},Y_{\rm cen}) = (-615\arcsec,-91\arcsec)$.
The raster scan has a temporal cadence of 9.6$\s$. The whole scan of the raster has 400 pixels in the $x$-direction with the pixel size 0.35$\arcsec$~and 1096 pixels in the $y$-direction with the pixel size 0.16$\arcsec$. The region of interest ranges from $-609\arcsec~{\rm to}~-559\arcsec$~in $x$-direction and $-79\arcsec~{\rm to}~-39\arcsec$~in $y$-direction (middle-column (panel E,F,G and H); Fig.~\ref{fig_mosaic1}).
Dataset 3: The dataset used for our study was observed by IRIS during the time
period from from 19:59:44 to 21:00:57 UTC on 29th March 2017 targeting AR\,12645.
IRIS observed ranges of  $141\arcsec$~in the $x$-direction and $175\arcsec$~in the
$y$-direction centred at ($X_{\rm cen},Y_{\rm cen}) = (-517\arcsec,-69\arcsec)$.
The raster scan has a temporal cadence of 9.2$\s$. The whole scan of the raster has 274.50 pixels in the $x$-direction with the pixel size 0.35$\arcsec$~and 200 pixels in the $y$-direction with the pixel size 0.35$\arcsec$. The region of interest ranges from $-529\arcsec~{\rm to}~-479\arcsec$~in $x$-direction and $-139\arcsec~{\rm to}~-91\arcsec$~in $y$-direction (right-column (panel I,J,K and L); Fig.~\ref{fig_mosaic1}).

\begin{figure*}
\centering
\mbox{
\includegraphics[trim=2.5cm 5.0cm -8.0cm -0.5cm,scale=0.45]{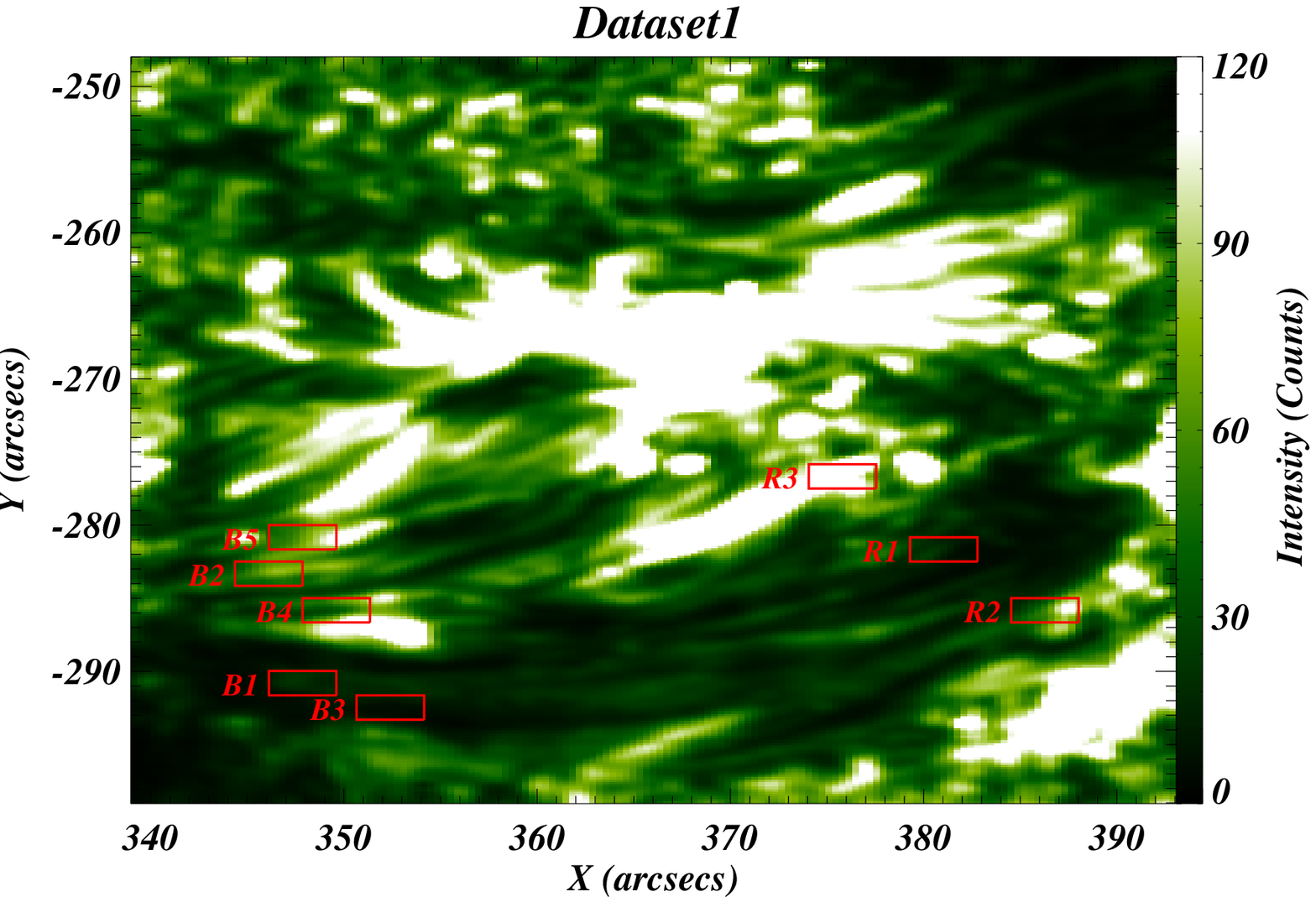}
\includegraphics[trim=8.0cm 0.5cm -1.0cm -1.0cm,scale=0.45]{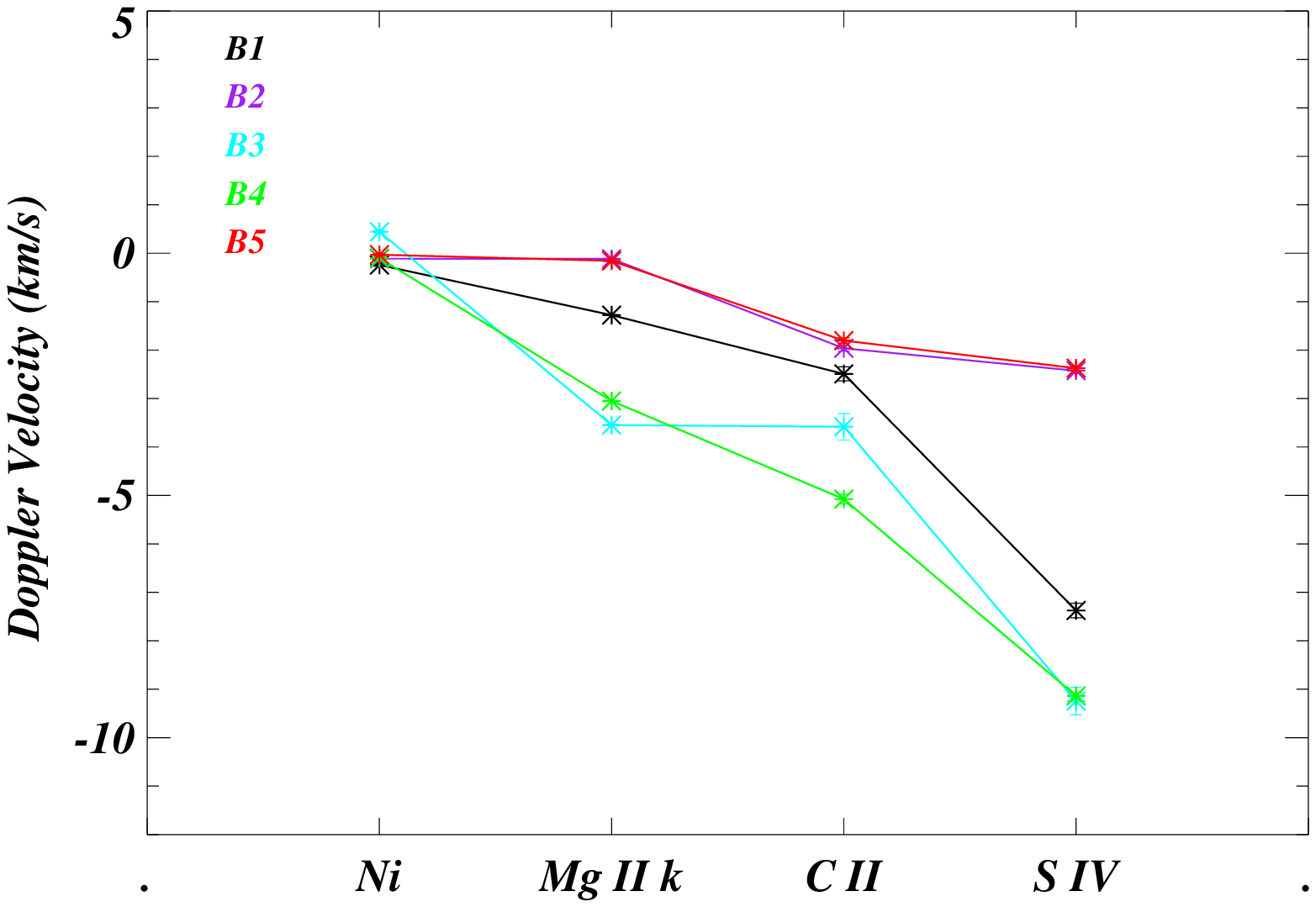} }
\mbox{   
\includegraphics[trim=-15.0cm 0.5cm -1.0cm 0.0cm,scale=0.45]{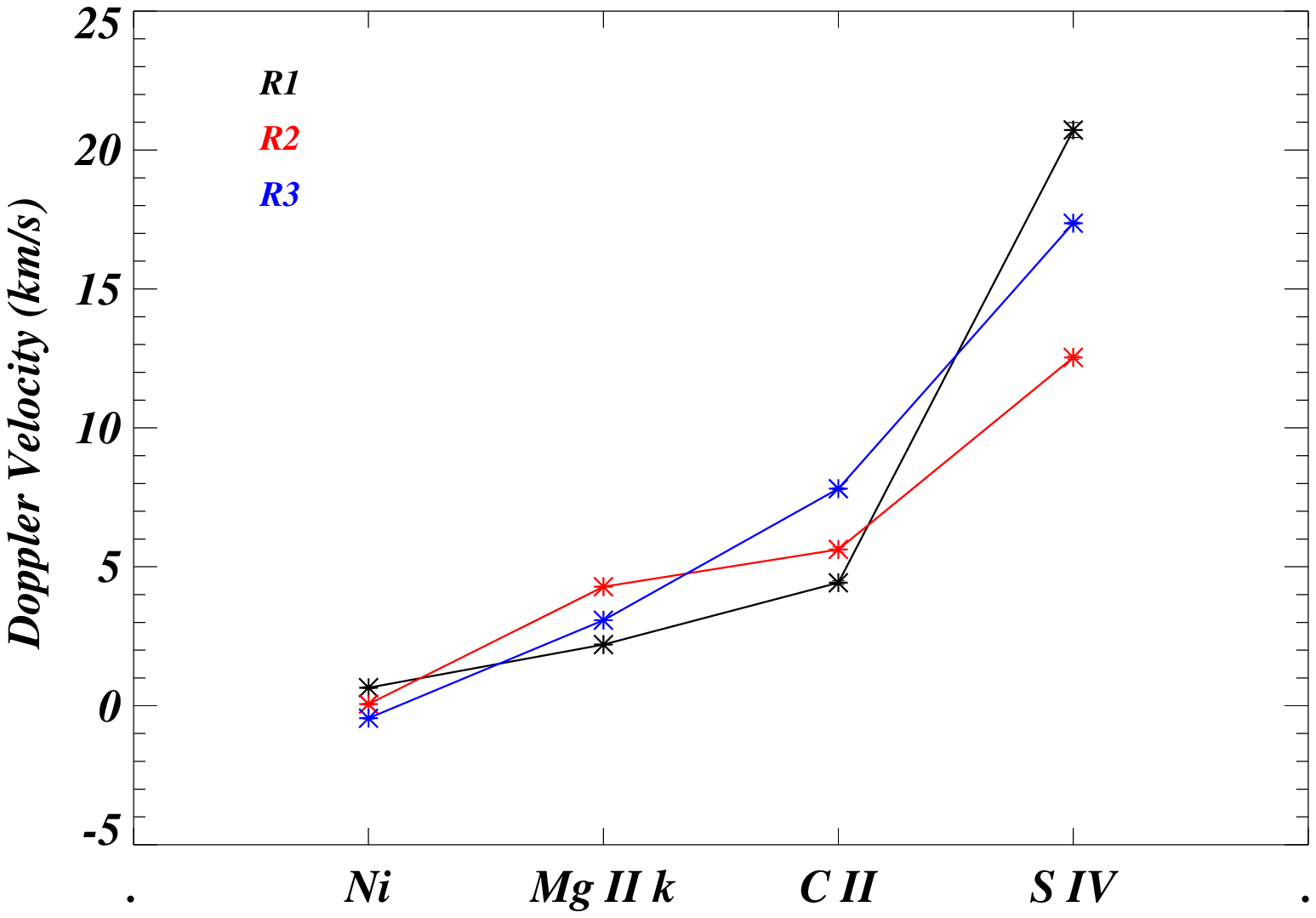}}     
\\
\caption{Left panel: The intensity map of Si\,{\sc IV} (1402.77~\AA) line with the boxes overlaid showing different locations at the footpoints of the cool loop systems. Top-right panel: The variation of Doppler velocity with the formation heights of different spectral lines for different boxes at the blueshifted footpoint indicated in the left panel. Bottom-right panel: The variation of Doppler velocity with the formation heights of different spectral lines for different boxes at the redshifted footpoint indicated in the left panel.}
\label{fig_vel1}

\end{figure*}

We analyze the spectroscopic observations of various lines formed over a broad range of temperatures. We have applied iris inbuilt routine (i.e., iris$\_$orbitvarr$\_$corr$\_$l2.pro) to remove the orbital variations before applying the fitting procedure. The estimation of rest wavelengths is very crucial for the estimation of Doppler velocities. We estimated the observed wavelengths of neutral lines in the very quiet area (Ni\,{\sc i} 2799.474~{\AA} and S\,{\sc i} 1401.50~{\AA}). Using the shift of the neutral lines, we have estimated the rest wavelengths of Ni\,{\sc i} 2799.474~{\AA}, Mg\,{\sc ii} k 2796.35 and Si\,{\sc iv} 1402.77~{\AA}. We estimate the observed wavelength of C\,{\sc ii} 1334.532~{\AA} in very quiet conditions, which is used as the rest wavelength of this line. The estimated rest wavelengths of these spectral lines in each dataset are tabulated in Table~\ref{table:table1}. 
To derive the basic parameters of the line profiles, we use Gaussian fitting to lines of Ni\,{\sc i}; formation temperature: $\log (T/\K)~=~4.2$ 
Stucki et al. (2000), 
Mg\,{\sc ii}\,k (2796.20~\AA; $\log (T/\K)$~=~4.0), C\,{\sc ii} (1334.53~\AA; $\log (T/\K)$~=~4.3), and Si\,{\sc iv} (1402.77~\AA; $\log (T/\K)$~=~4.8) (cf., example fits in Fig.~\ref{fig_spec1}).
The temperature coverage for different lines is indicated as provided by De Pontieu et al. (2014).

We have performed single Gaussian fitting to the Si\,{\sc iv} line, inverse Gaussian fitting to the Ni\,{\sc i}  lines, and single or double Gaussian fitting, depending on the nature of observed profiles, to Mg\,{\sc ii}\,k and C\,{\sc ii} lines. Since Mg\,{\sc ii}\,k lines are optically thick, they provide double Gaussian profiles at most of the locations except for sunspots (Morrill et al. 2001; Leenaarts et al. 2013; Rathore et al. 2015). The details of fitting the optically thick lines are provided in Appendix~\ref{section:append}. 

\begin{table*}
\begin{center}
\begin{tabular}{|c|c|c|c|c|}
\hline
 .           & Ni\,{\sc I}~(\AA)        &  Mg\,{\sc ii}\,k3~(\AA)     & C\,{\sc ii}~(\AA)          & Si\,{\sc iv}~(\AA)\\
\hline
Dataset 1   	& 2799.4745     &  2796.3512       	    & 1334.5391             & 1402.7770 \\
Dataset 2    	& 2799.4731     &  2796.3498            & 1334.5374             & 1402.7940 \\
Dataset 3     	& 2799.4777     &  2796.3544            & 1334.5368             & 1402.8001 \\
\hline
\end{tabular}
\end{center}
\caption{Calibrated wavelengths for different spectral line for different datasets}
\label{table:table1}
\end{table*}
\section{Observational results and their interpretation} \label{sec:results}
\subsection{Dataset 1} 
\label{ss.1}
\textbf{Identification of the cool loop system:}
The dynamic loop structures, which are typically visible at low temperatures of the chromosphere and TR, correspond to the cool loops. These cool loop system are present in all three datasets. The left-column of Fig.~\ref{fig_mosaic1} (i.e., panel A,B,C and D) shows the mosaic of intensity images of different spectral lines taken from the IRIS raster data along with the LOS magnetogram taken from HMI/SDO to show the cool loop system and its footpoint magnetic polarities. To create these intensity images, we have taken the averaged emission within the certain wavelengths around the central wavelengths of these lines, e.g., Mg\,{\sc ii} (2795.0 to 2797.2)~{\AA}, C\,{\sc ii} (1334.0 to 1335.0)~{\AA} and Si\,{\sc iv} (1402.0 to 1404)~{\AA}. This system has been taken into consideration as Dataset 1. The LOS magnetogram (Fig.~\ref{fig_mosaic1}; panel A) shows the presence of opposite magnetic polarities in which the foot points of the cool loop system lie. This is multi-threaded and dynamic loop system as observed by IRIS. In the present study, we use multiple spectral lines covering wide temperature range to understand Doppler flows pattern and its variation  in the observed cool loop system.
The Mg\,{\sc ii}\,k line shows doubly peaked profiles and exhibits emission reversal for all the regions except in the sunspots (Morrill et al. 2001). The Mg\,{\sc ii}\,k (panel B; Fig.~\ref{fig_mosaic1}) and C\,{\sc ii} (Fig.~\ref{fig_mosaic1}; panel C) lines correspond to the chromospheric plasmas, which show the evidence of the cool loops. Although, the top part of the loop system is not fully evident in some cool lines, however, its lower part above the footpoints are clearly visible. The loop system along with the different loop strands is clearly identified in the Si\,{\sc iv} (Fig.~\ref{fig_mosaic1}; panel D) line along with its footpoint at both the ends. \\
\begin{figure*}
\mbox{
\hspace{-0.5cm}
\includegraphics[trim=2.5cm 0.5cm 0.2cm 0.2cm,scale=0.6]{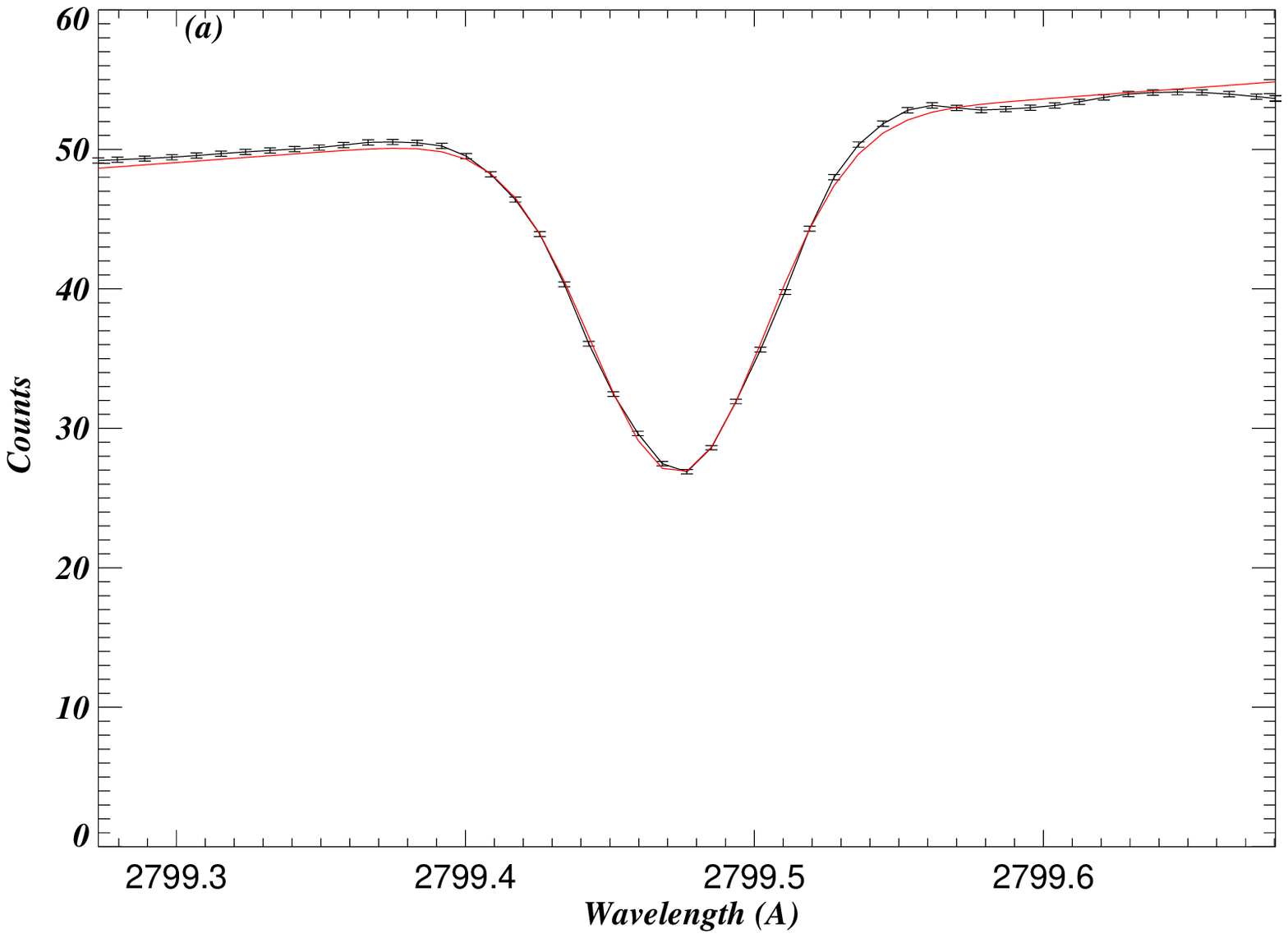}
\includegraphics[trim=1.0cm 0.5cm 0.2cm 0.2cm,scale=0.6]{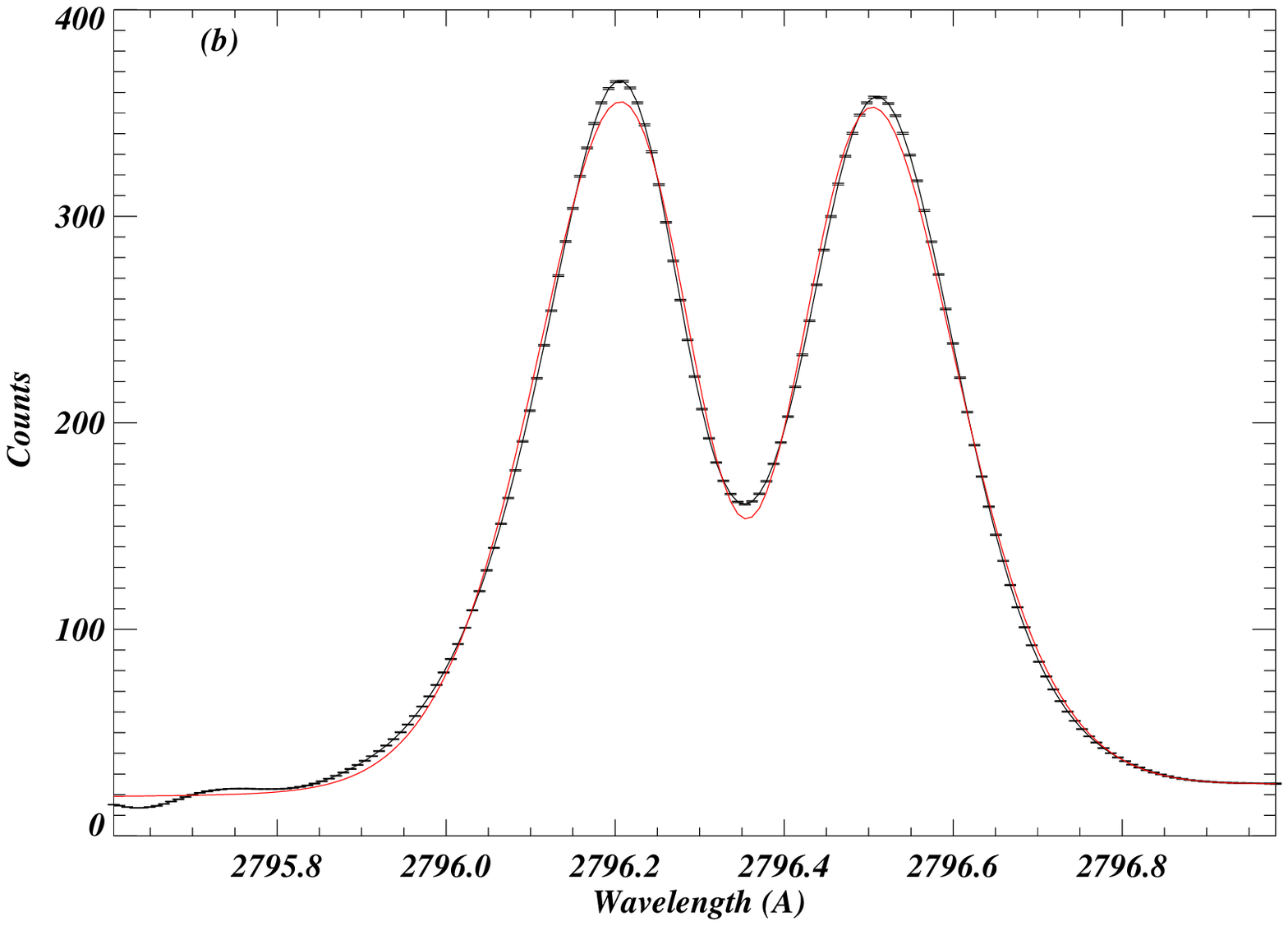}}
\mbox{
\hspace{-0.5cm}
\includegraphics[trim=2.5cm 0.5cm 0.2cm 0.2cm,scale=0.6]{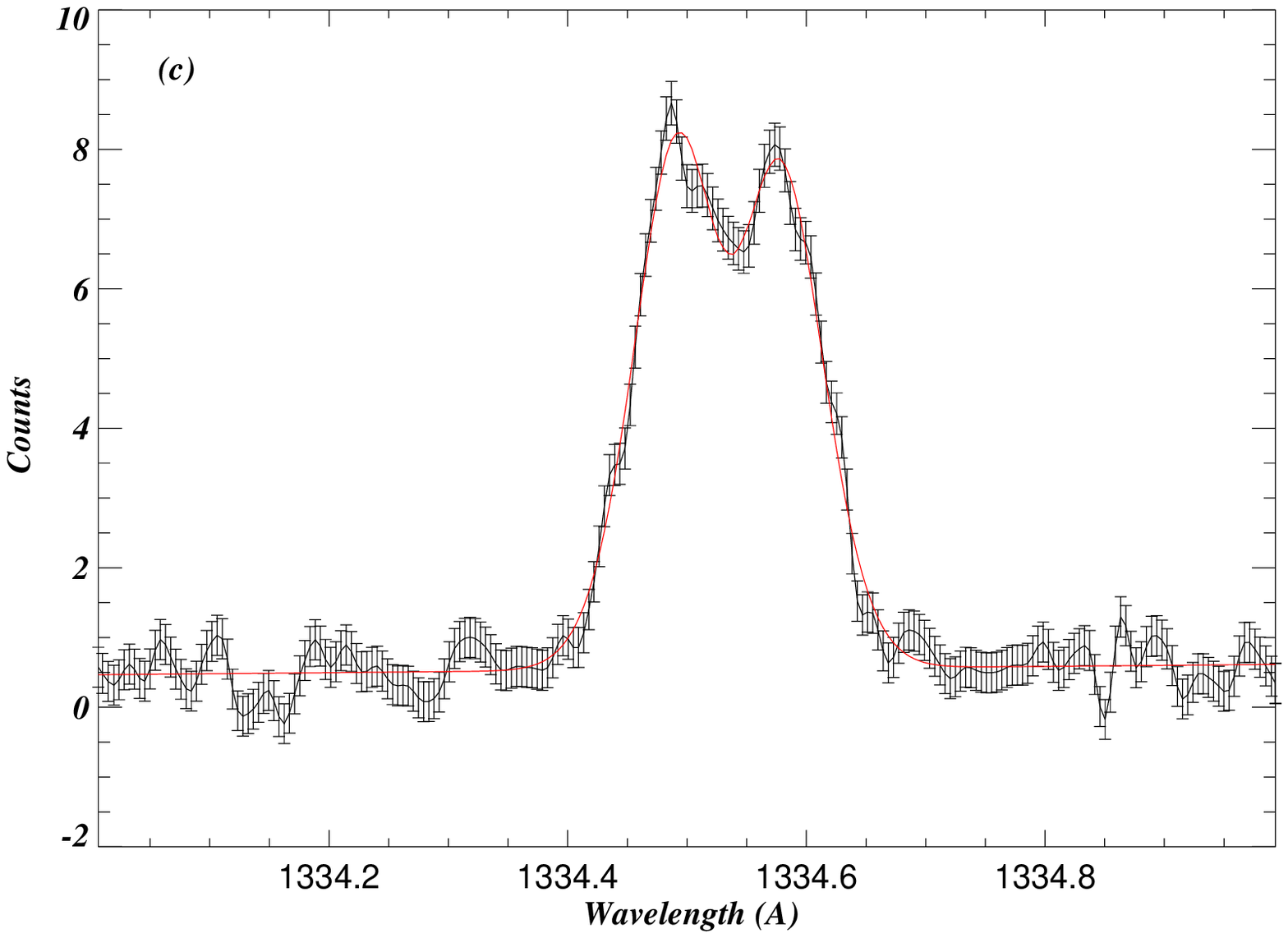}
\includegraphics[trim=1.0cm 0.5cm 0.2cm 0.2cm,scale=0.6]{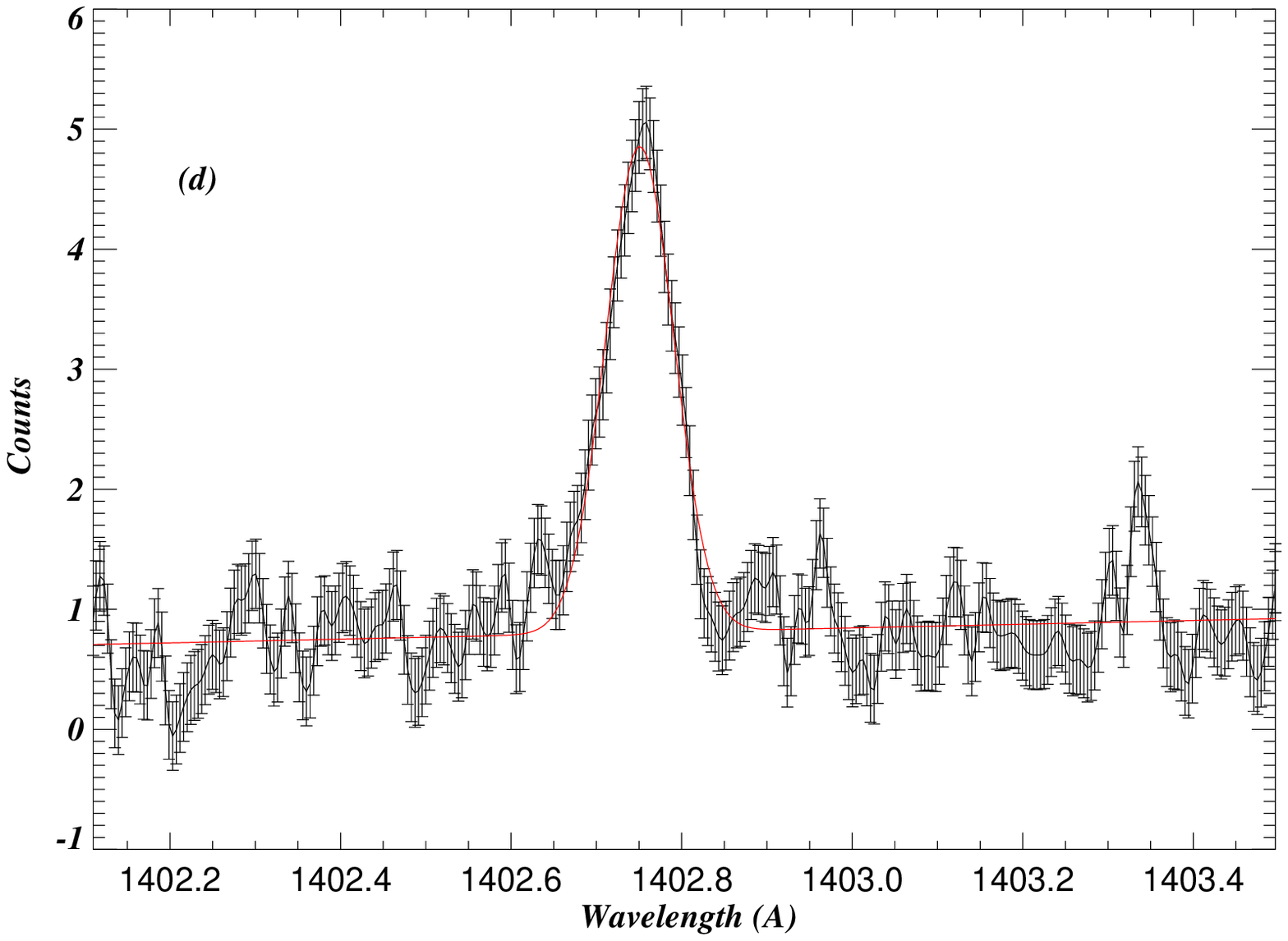}
}
\caption{Spectral fitting of (a) Ni\,{\sc I} (2799.47~\AA), (b) Mg\,{\sc II}\,k (2796.2~\AA), (c) C II (1354.53~\AA), and (d) Si\,{\sc IV} (1402.77~\AA) lines for averaged profile over the box labelled as B4 shown in the left panel of Fig.~\ref{fig_vel2} corresponding to Dataset 2.}
\label{fig_spec2}
\end{figure*}


\textbf{Various parametric maps for the first dataset:}
We have selected various boxes near the footpoints of these cool loops, which are shown in the left panel of Fig.~\ref{fig_vel1} at Si\,{\sc IV} intensity map. Some boxes are located in blue-shifted end (e.g., B1,B2, etc) while others are located in the red-shifted end (e.g., R1,R2, etc.). These locations of the boxes are chosen to have higher emissions such that the intensity along the strand becomes more than double at the footpoint. The averaged spectra for each box is obtained by averaging all existed spectra within the particular box. Similar criteria for the selection of boxes has been followed by other datasets (Dataset 2 and Dataset 3). Fig.~\ref{fig_spec1} presents averaged profiles of different spectral lines used in our analysis. The spectral profiles correspond to the box
labelled as B5 for Dataset 1 (Fig.~\ref{fig_vel1}). The black solid line represents the averaged spectral profile for different spectral lines where corresponding errors are indicated by the error bars. The red solid line is the fitted profile on the observed profiles (black lines)
We have fitted the each observed profiles within ROI to get the intensity, Dopper velocity and FWHM maps.
Left column of Fig.~\ref{fig_data1} shows the intensity (top-left), Doppler velocity (middle-left) and FWHM maps (bottom-left) of the Mg\,{\sc ii} k. In the similar fashion, we have shown the intensity, Doppler velocity and FWHM maps of C\,{\sc ii} 1334.53~{\AA} (middle-column) and Si\,{\sc iv} 1403.53~{\AA} (right-column) in Fig.~\ref{fig_data1}. 
Since Mg\,{\sc ii}\,k is an optically thick line, it shows doubly peaked profile which has been fitted by using two Gaussians (positive and negative).
Similar properties hold for the optically thick line C\,{\sc ii} 1334.53~{\AA}. 
For Mg\,{\sc ii} 2796.35~{\AA} and C\,{\sc ii} 1334.53~{\AA} lines, we have taken the parameters from negative Gaussian, which captures the minima between two peaks. This minima basically forms in the upper chromosphere/TR. For the location exhibiting the single peak, we have taken the same parameters from positive Gaussian because only one Gaussian is present in this case. The detailed description is provided in appendix~\ref{section:append}. Using this methodology, we have produced the intensity, Doppler velocity and FWHM maps for Mg\,{\sc ii} k and C\,{\sc ii} lines. The estimation of these maps from Si\,{\sc iv} line is straightforward as the line is single peak line.
For Mg\,{\sc ii}\,k, the intensity map (top-left panel; Fig.~\ref{fig_data1}) shows the loop structures for which the footpoints are not distinct but the top part is clearly visible. Doppler velocity map (middle-left panel; Fig.~\ref{fig_data1}) shows blue-shifted velocities at one end and red-shifted at another which has been mapped having values ranging from (-10.0~{\rm to}~+10.0) $\km\s^{-1}$. The corresponding FWHM map is shown in the bottom-left panel of Fig.~\ref{fig_data1}. The line width range for Mg\,{\sc ii} k varies from (0.10 to 0.35)~{\AA}.
\begin{figure*}
\mbox{
\hspace{-1.0cm}
\includegraphics[trim=6.0cm 1.0cm 5.0cm 0.5cm,scale=0.9]{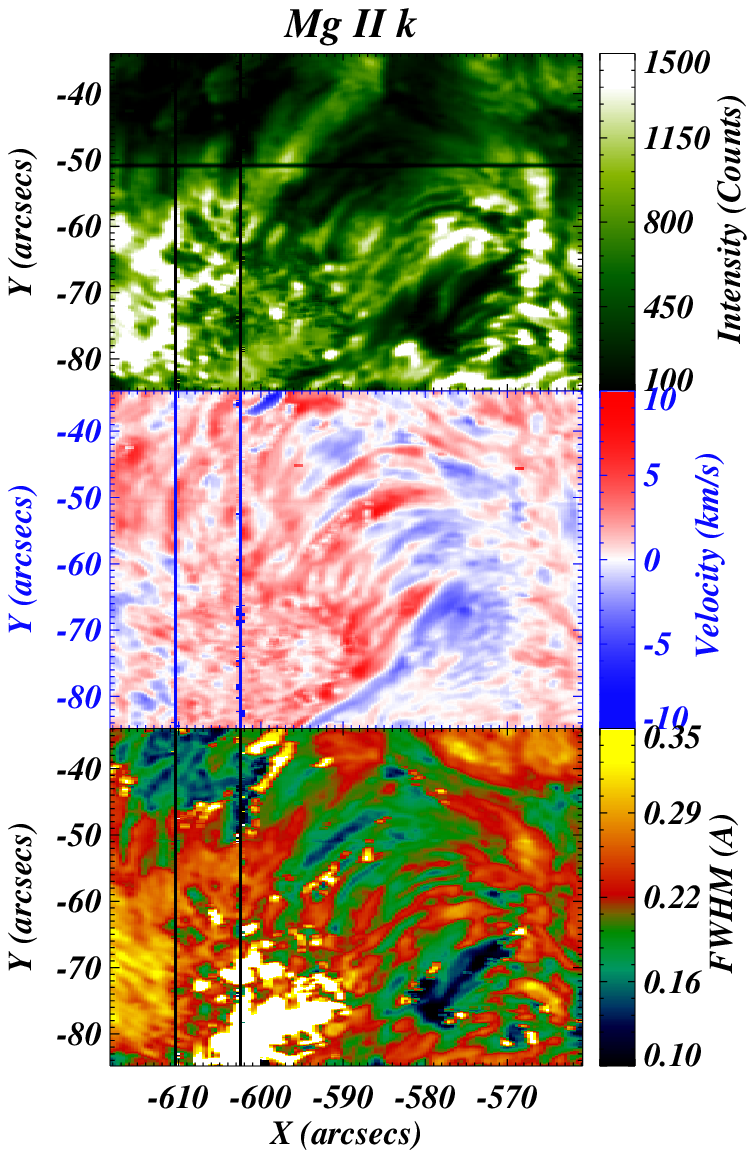}
\includegraphics[trim=5.0cm 1.0cm 5.0cm 0.5cm,scale=0.9]{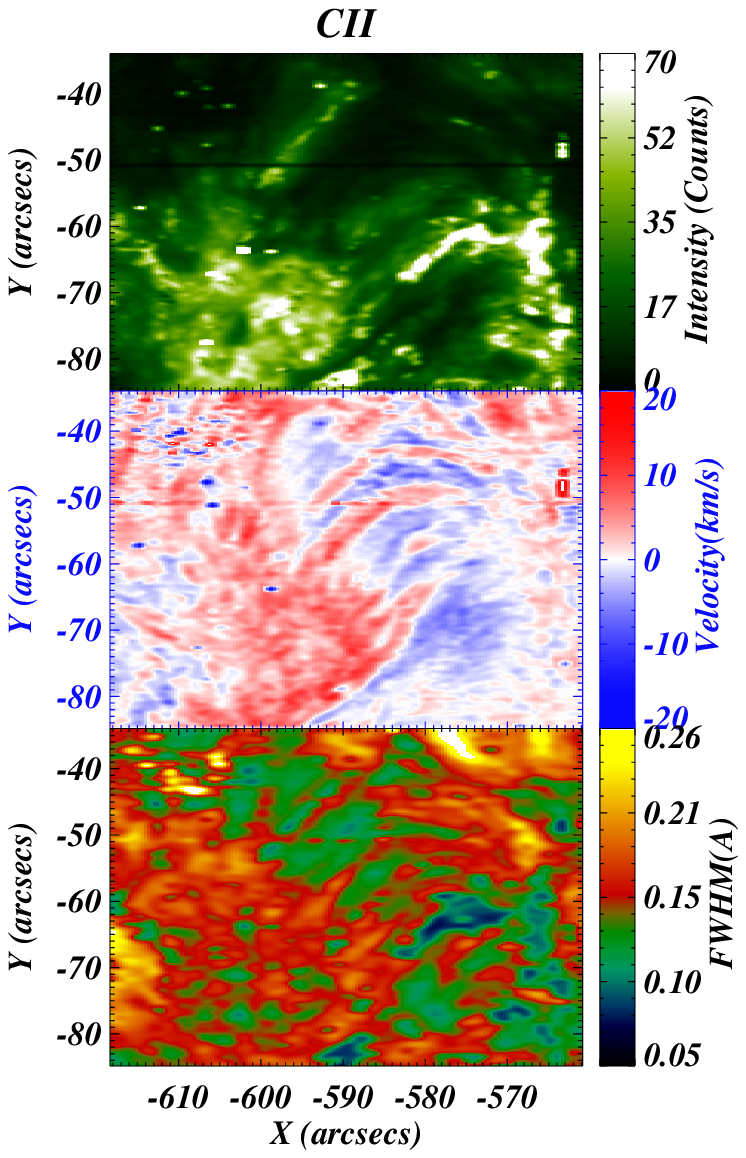}
\includegraphics[trim=5.0cm 1.0cm 5.0cm 0.5cm,scale=0.9]{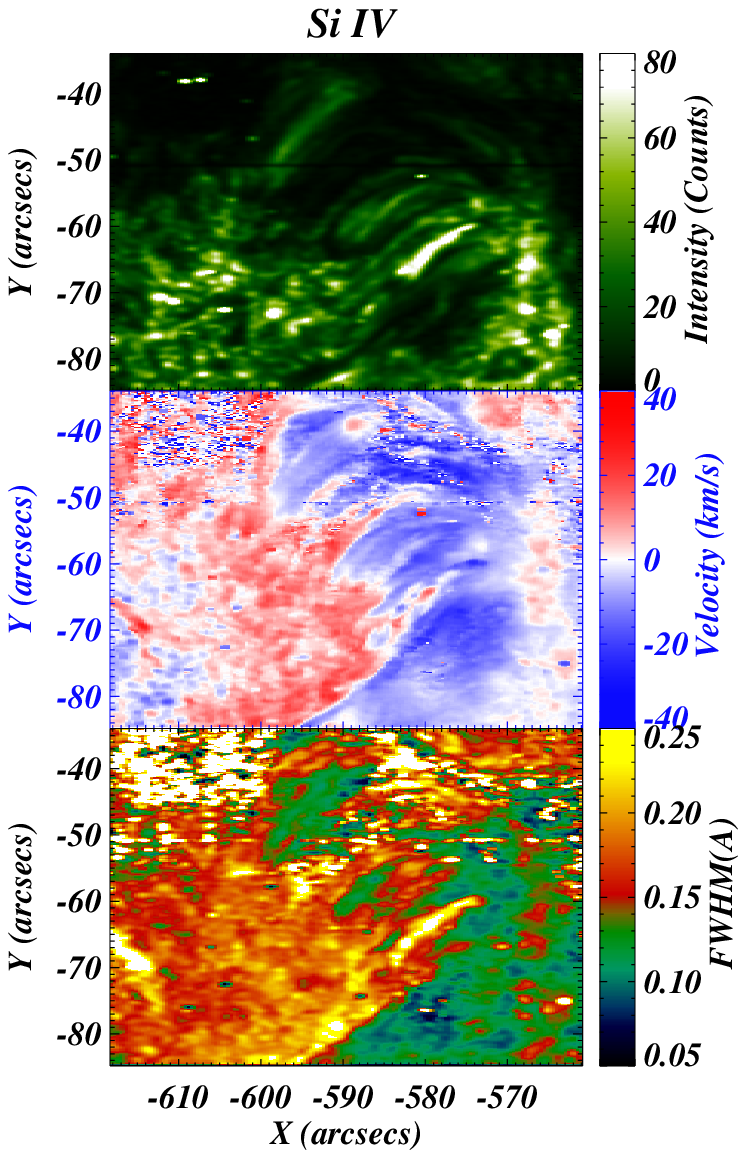}
}
\caption{Intensity, Doppler velocity, and Full width at half maximum (FWHM) maps
of Mg\,{\sc II}\,k (2796.2~\AA) ,
C\,{\sc II} (1334.53~\AA), and
Si\,{\sc IV} (1402.77~\AA)
lines are shown for the Dataset 2
in the left, middle, and right columns respectively.}
\label{fig_data2}
\end{figure*}


\begin{figure*}[h]
\centering
\mbox{
\includegraphics[trim=2.5cm 5.0cm -8.0cm -0.5cm,scale=0.45]{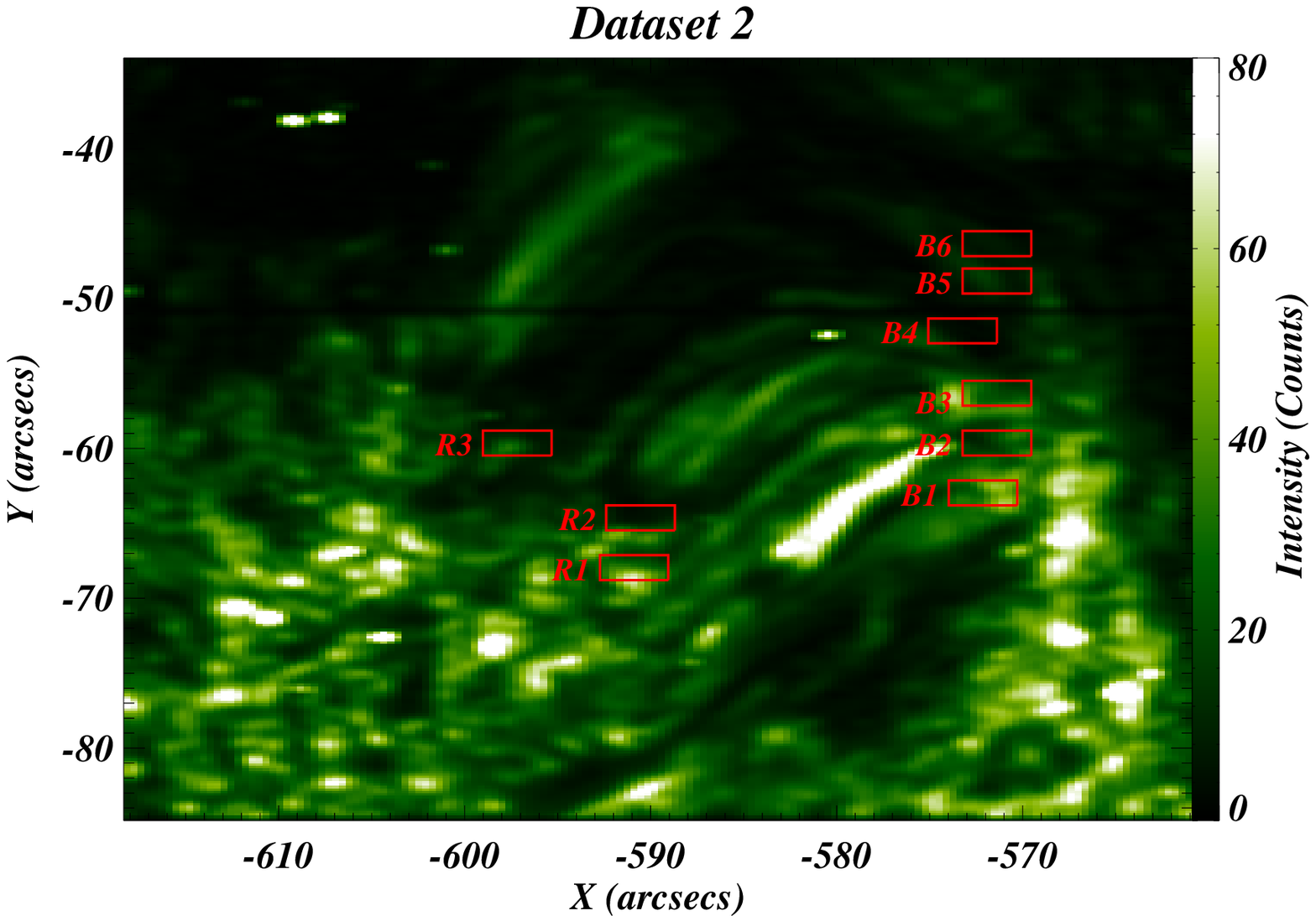}
\includegraphics[trim=8.0cm 0.5cm -1.0cm -1.0cm,scale=0.45]{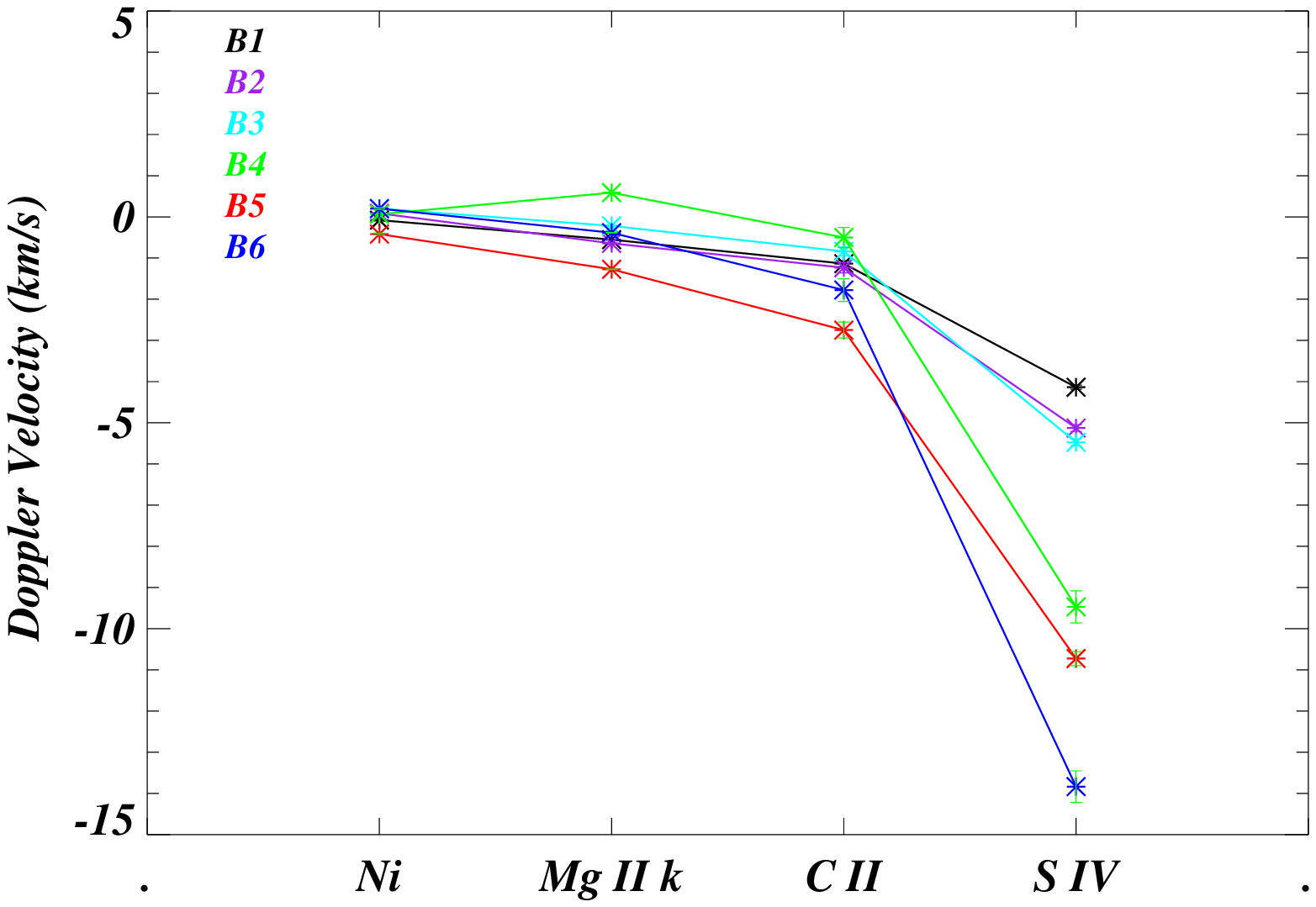} }
\mbox{   
\includegraphics[trim=-15.0cm 0.5cm -1.0cm 0.0cm,scale=0.45]{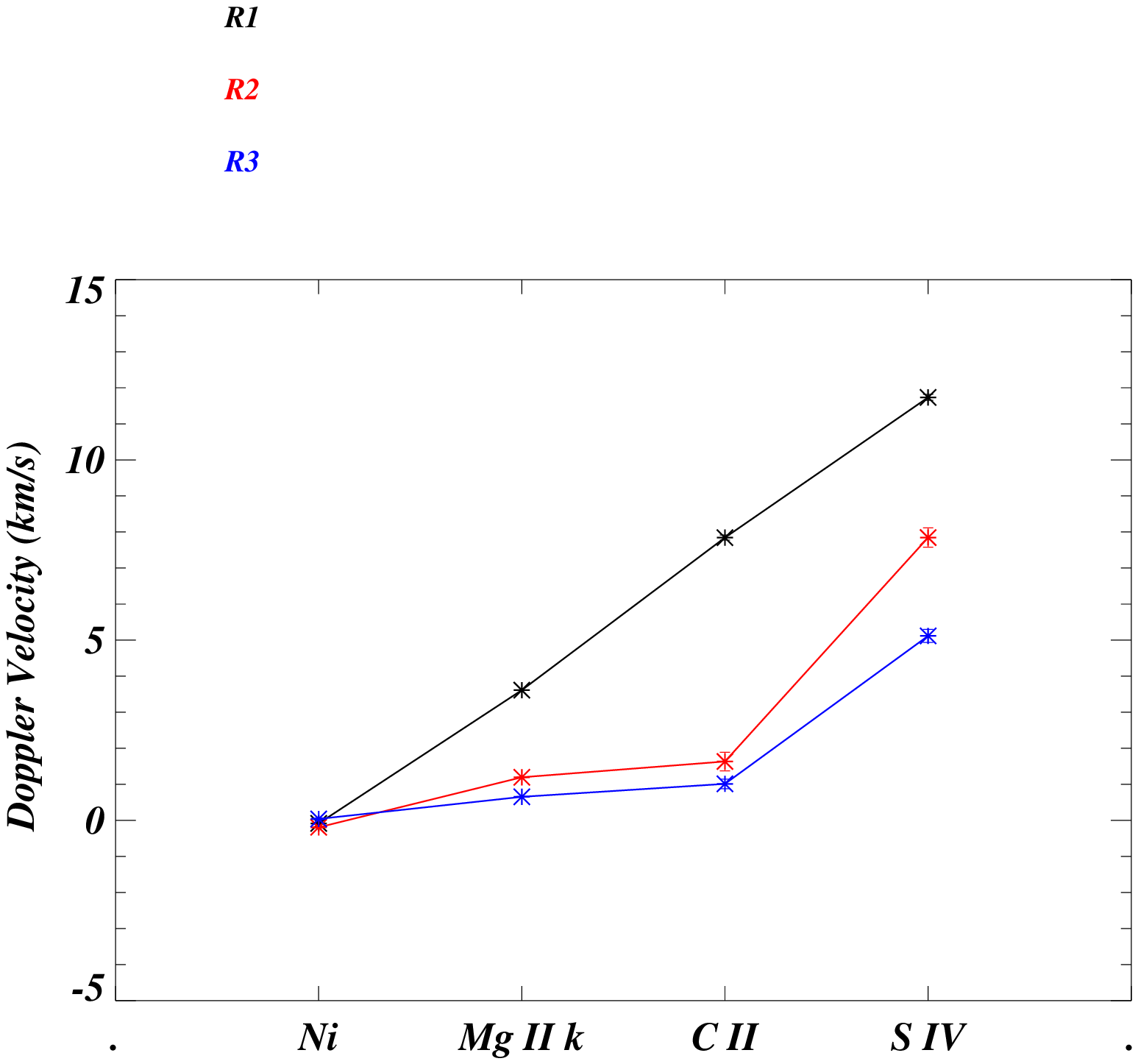}}     
\\
     
\caption{Left panel: The intensity map of Si\,{\sc IV} (1402.77~\AA) line with the boxes overlaid showing different locations at the footpoints of the cool loop systems. Top-right panel: The variation of Doppler velocity with the formation heights of different spectral lines for different boxes at the blueshifted footpoint indicated in the left panel. Bottom-right panel: The variation of Doppler velocity with the formation heights of different spectral lines for different boxes at the redshifted footpoint indicated in the left panel.}
\label{fig_vel2}

\end{figure*}

For C\,{\sc ii} line, such parametric maps are shown in the middle-column of Fig.~\ref{fig_data1} where the fuzzier loops are visible (Top panel; Fig.~\ref{fig_data1}) along with the direction of plasma flow indicated by Doppler velocity map (Middle panel; Fig.~\ref{fig_data1}(ii)). The FWHM map of C\,{\sc ii} 1334.53~{\AA} is displayed in bottom panel of Fig.~\ref{fig_data1}(ii), which has values ranging from 0.05~\AA~ to 0.30~\AA~. The cool loop system for Si\,{\sc iv} line having
different strands is clearly visible in the intensity image (Top panel; Fig.~\ref{fig_data1}) 
representing plasma maintained at TR temperature.
The corresponding Doppler velocity shows the blueshifted and redshifted opposite footpoints of the cool loop system (Middle panel; Fig.~\ref{fig_data1}). The Si\,{\sc iv} Doppler velocity map has also been discussed by Huang et al. (2015) targeting the same AR using Si\,{\sc iv} line with different time duration (21:02 UT to 21:36 UTC) corresponding to different dataset.
Their results are qualitatively similar to what we have observed in the Doppler velocity 
map of Si\,{\sc iv} line. We have then correlated Doppler velocity maps with FWHM maps for  Si\,{\sc iv} line as well as some other lines which have been shown in Fig.~\ref{fig_data1}.
One end of the cool loop system is dominated by redshifts (downflows), while the other end is dominated by blueshifts (upflows). This change from blueshifts at one end to the redshifts at the other end shows the direction of plasma flows. The plasma flow structures are similar to the loop structure visible in the intensity map of the Si\,{\sc iv} line (Top panel; Fig.~\ref{fig_data1}). The blueshifted region shows lower values of FWHM and the redshifted end shows higher FWHM values (Bottom panels; Fig.~\ref{fig_data1}). The corresponding values of FWHM ranging from (0.05~\AA~ to 0.25~\AA~) are shown in the colorbar which exhibit the characteristics of TR. The increased linewidth at redshifted (downflows) footpoint may be due to the heating caused by downflowing mass motions there, which are attributed to the line broadening (Tian et al. 2008; Tian et al. 2009).

\textbf{Variation of Doppler velocity:}
The boxes overlaid on the intensity map of Si\,{\sc iv} line are labelled as B1, B2, B3, B4, B5 and R1, R2, R3 (Left-panel; Fig.~\ref{fig_vel1}). These boxes indicate the locations of the footpoints of different loop strands chosen to investigate the Doppler velocities above the footpoints of the cool loop threads. The average spectral profiles from these boxes are used to infer qualitatively the Doppler velocity pattern in these regions.
The photospheric signature of the footpoints of the loop system in SDO/HMI having positive and negative polarities at the opposite ends shows almost bipolar magnetic loops. Assuming the loops to have similar behaviour at the footpoints, plasma flows are investigated using Doppler velocities of multi-spectral lines. The variation of Doppler velocities is shown in the top-right panel of Fig.~\ref{fig_vel1} corresponding to the five different boxes (i.e., B1, B2, B3, B4, and B5) at the blue-shifted footpoints of cool loop systems as shown in left panel of Fig.~\ref{fig_vel1}. 
The Doppler velocity of the Ni\,{\sc i} line is very low in all boxes showing almost no upflows or downflows (-0.24 {\rm to} 0.44) $\km\s^{-1}$ in the photospheric region.
These blueshifts (upflows) increase for B1, B3, and B4 while they remain nearly
same for B2 and B5 up to the formation temperature of Mg\,{\sc ii}\,k. These boxes at formation temperature of Mg\,{\sc ii} k lines show considerable blue-shifts (-0.11 {\rm to} -3.54) $\km\s^{-1}$. However, the significant change in the blueshifts occurs at the formation temperature of C\,{\sc ii} for the different boxes as revealed by the top-right panel of Fig.~\ref{fig_vel1}. The Doppler velocity variation at Si\,{\sc iv} shows blueshifts having -2.42$\km\s^{-1}$  and  -2.37$\km\s^{-1}$ value for B2 and B5 while
the B1, B3 and B4 have high blueshifts (i.e., -7.37$\km\s^{-1}$--B1, -9.24$\km\s^{-1}$--B2 and -9.14$\km\s^{-1}$--B4). 
\\We have also considered three location labelled as R1, R2, and R3  in the redshifted region within the other footpoints of this cluster of cool loops. The Doppler velocities increase for different spectral lines showing the increase in plasma downflows as we go
higher up in the solar atmosphere as shown in the bottom-right panel of Fig.~\ref{fig_vel1}. 
The Doppler velocity for Ni\,{\sc i} line shows small flows having range (-0.44 {\rm to} 0.64) $\km\s^{-1}$ for different locations.
The optically thick lines (Mg\,{\sc ii} k and C\,{\sc ii}) are fitted as discussed in appendix \ref{section:append}. The centroid of the negative Gaussian is used to estimate the corresponding Doppler velocities which ranges from middle chromosphere to the upper chromosphere.
Mg\,{\sc ii}\,k line has significant downflows having Doppler velocity range of (2.20 {\rm to} 4.42) $\km\s^{-1}$. The downflows are then increased as we go higher to the formation height of C\,{\sc ii} having Doppler velocity range of (4.28 {\rm to} 7.81) $\km\s^{-1}$ These values reach to the maximum values of (12.53 {\rm to} 20.72) $\km\s^{-1}$ for Si\,{\sc iv} line which is formed in the transition region. The maximum activity is found in the upper chromosphere/TR interface where the downflows
become large and create excess widths at the associated footpoints. The corresponding 1-sigma error in the line centroid of the fitted profile converted into velocity terms has been shown in terms of error bars for different lines at different positions. The 1-sigmaa velocity error being very small, it is hard to visualize these errors in the corresponding figures (cf., Fig.~\ref{fig_vel1}).

\textbf{Line widths of Si\,{\sc iv} at red and blue-shifted footpoints of cool loops:}
We have also investigated the FWHM of the Si\,{\sc iv} line within all boxes (blue as well as red) for the first data set. The FWHM of the blue boxes of the first observation is 0.095~{\AA} (B1), 0.127~{\AA} (B2), 0.114~{\AA} (B3), 0.153~{\AA} (B4), and 0.123~{\AA} (B5). While, the red boxes have widths of 0.215~{\AA} (R1), 0.185~{\AA} (R2), and 0.246~{\AA} (R3). 

Thermal width ($\sigma_{th}$) of Si\,{\sc iv} line is 0.09~{\AA} while the instrumental width ($\sigma_{inst}$) is 0.026~{\AA} (De Pontieu et
al. 2014). Therefore, the non-thermal component of the line-widths 
($\sigma_{nt}$=[$\sigma_{obs}^{2}$-$\sigma_{th}^{2}$-$\sigma_{inst}^{2}$]$^{1/2}$)
derived by the averaged spectral line profiles at different blue-shifted boxes (B1 - B5) range between 0.015~{\AA} to 0.120~{\AA}. The plasma upflow is also associated with these regions of increased non-thermal width. In the present observational base-line, we conjecture the presence of localized energy release due to small-scale magnetic reconnection above the footpoints of the cool loop threads, and filling-up of the evaporated plasma in these threads which we observe in form of the blue-shift (e.g., Patsourakos \& Klimchuk 2006; Hansteen et al. 2014; Huang et al. 2015; Polito et al. 2015). 

It is obvious that the red-shifted footpoints have higher line widths (almost double) compared to the blue-shifted footpoints. 
The downfall or condensation of the cool-loop plasma towards their red-shifted foot-points in TR may contribute into the line-width broadening
(Tian et al. 2009).
There may be additional physical processes which may cause excess line broadening on the red-shifted TR footpoints of the loop system, {\it viz.}, nano-flare generated acoustic waves (Hansteen 1993) from the overlying atmosphere, over-all down-ward propagating pressure disturbances (Zacharias et al. 2018), etc.

\subsection{Dataset 2}\label{ss.2}
\textbf{Identification of cool loops:}
The middle-column of Fig.~\ref{fig_mosaic1} shows the region of interest consisting of cool loop system for Dataset 2 along with the magnetic polarities at the footpoints shown by LOS magnetogram (Fig.~\ref{fig_mosaic1}; panel E).
The observational signatures of the existence of cool loop system are shown using Mg\,{\sc ii}\,k (Fig.~\ref{fig_mosaic1}; panel F), C\,{\sc ii} (Fig.~\ref{fig_mosaic1}; panel G) and  Si\,{\sc iv} (Fig.~\ref{fig_mosaic1}; panel H) lines intensity maps.
The intensity map of Si\,{\sc IV} line (Fig.~\ref{fig_mosaic1}; panel H) shows faint loops corresponding to the temperatures of the transition region. The characteristic behaviour of the loop system for different spectral lines is similar to that of description for Dataset 1 (left-column; Fig.~\ref{fig_mosaic1}).

\begin{figure*}
\mbox{
\hspace{-0.5cm}
\includegraphics[trim=2.5cm 0.5cm 0.2cm 0.2cm,scale=0.6]{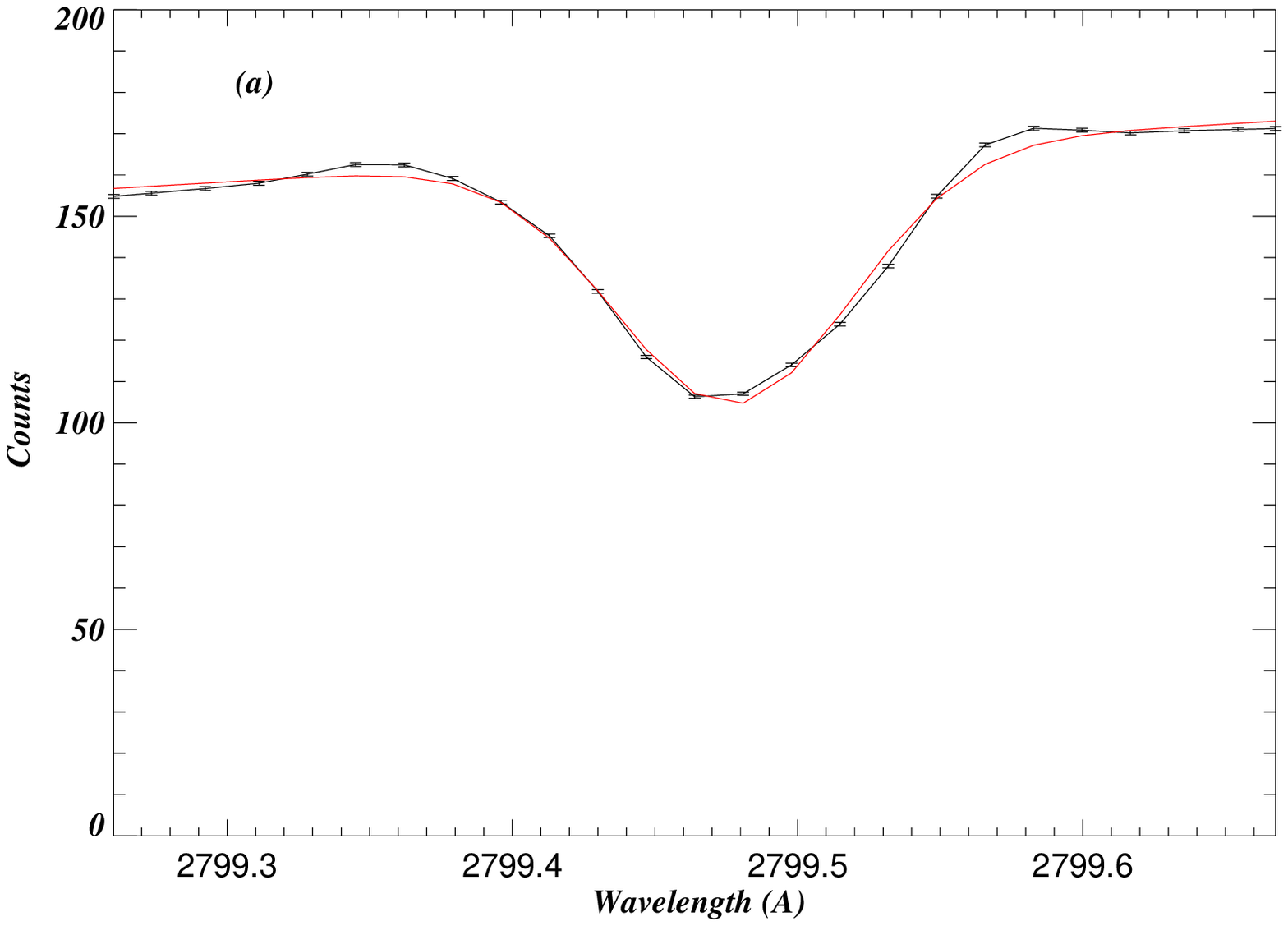}
\includegraphics[trim=1.0cm 0.5cm 0.2cm 0.2cm,scale=0.6]{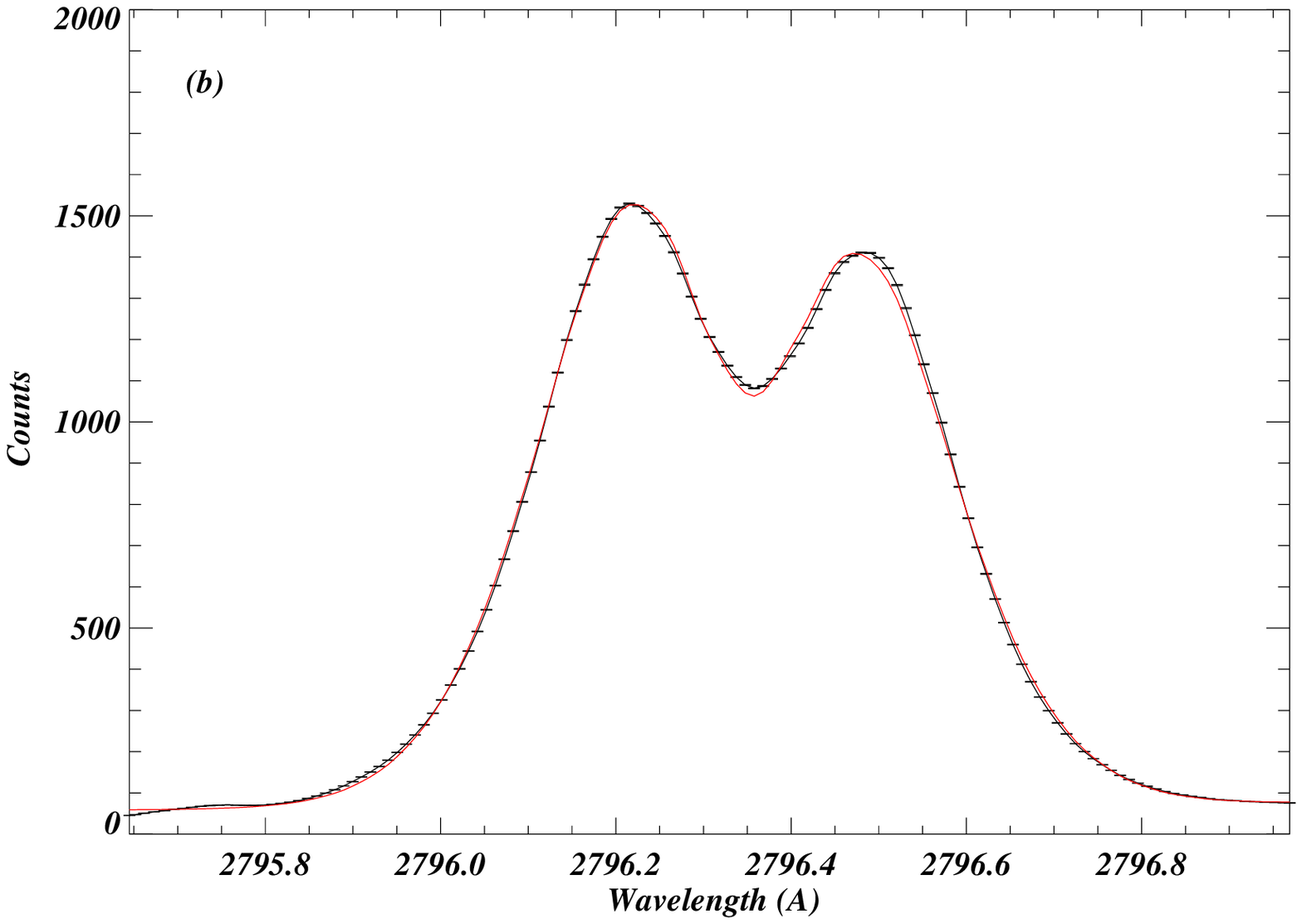}}
\mbox{
\hspace{-0.5cm}
\includegraphics[trim=2.5cm 0.5cm 0.2cm 0.2cm,scale=0.6]{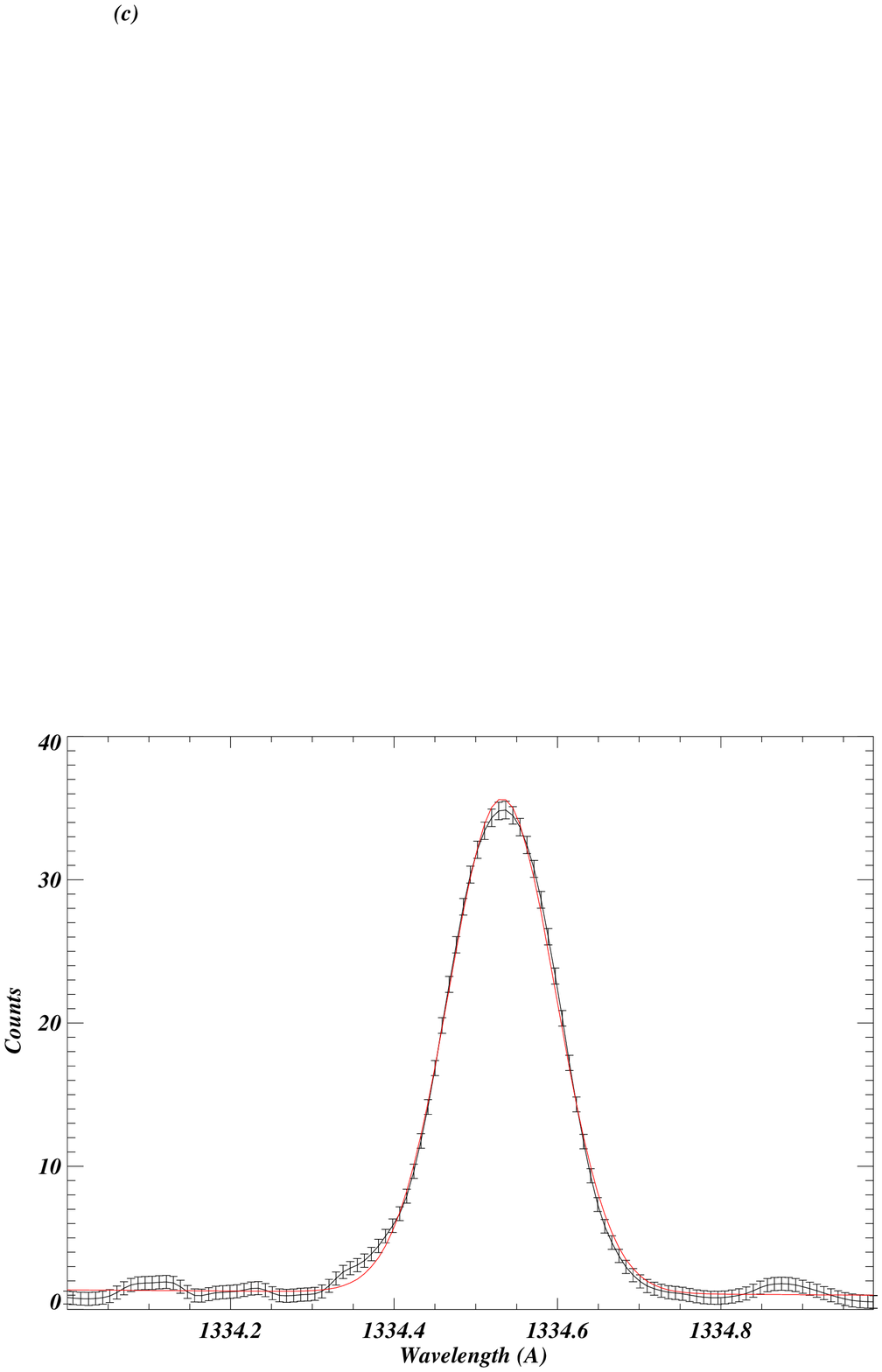}
\includegraphics[trim=1.0cm 0.5cm 0.2cm 0.2cm,scale=0.6]{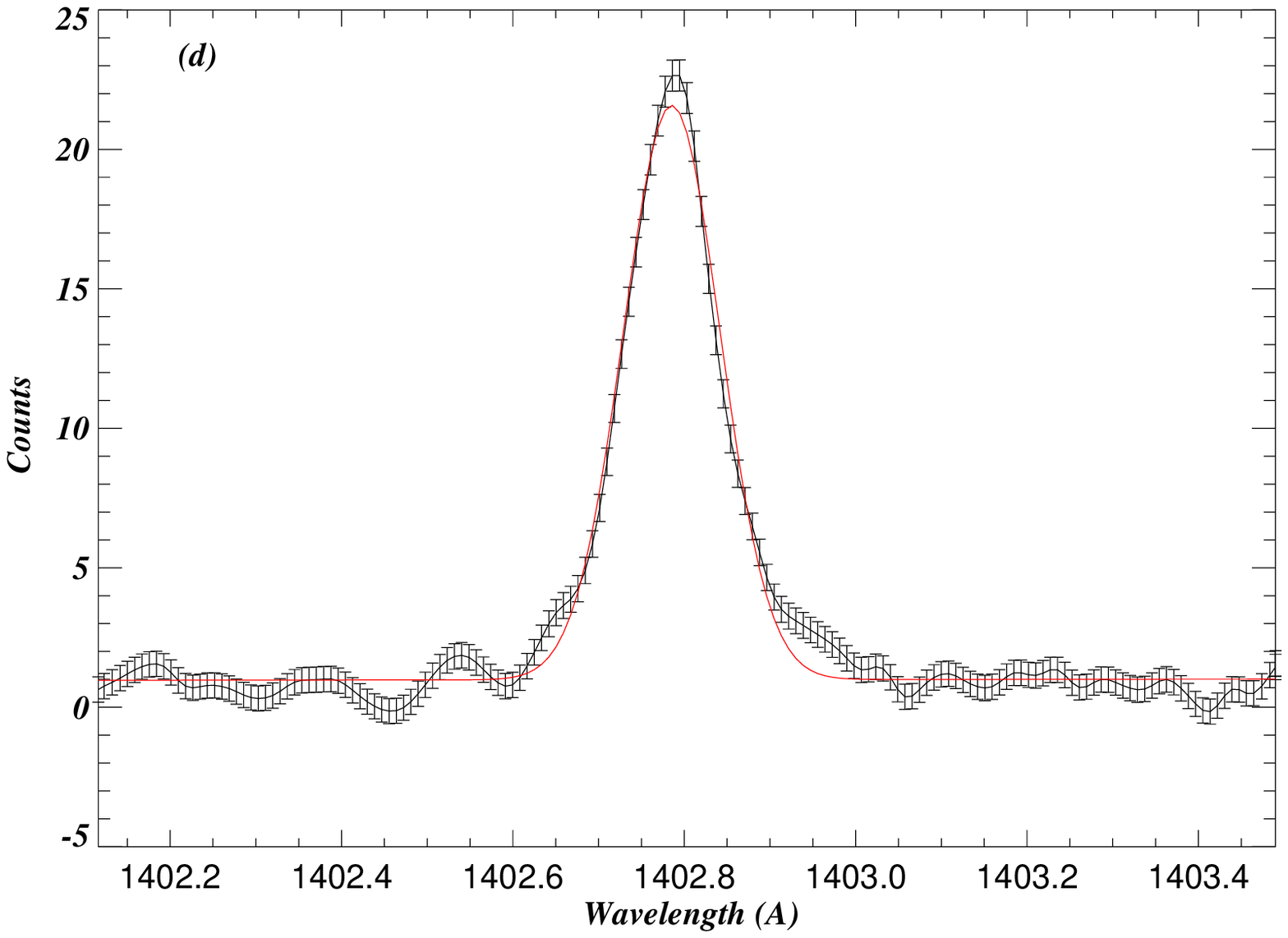}
}
\caption{Spectral fitting of (a) Ni\,{\sc I} (2799.47~\AA), (b) Mg\,{\sc II} k (2796.2~\AA), (c) C II (1354.53~\AA), and (d) Si\,{\sc IV} (1402.77~\AA) lines for box labelled as B3 in the left panel Fig.~\ref{fig_vel3} for Dataset 3.}
\label{fig_spec3}
\end{figure*}
\textbf{Parametric maps for the second dataset:}
Fig.~\ref{fig_spec2} represents average observed profiles (averaged over the selected boxes B4) of different spectral lines. The displayed spectral profiles correspond to the box labelled as B4 for Dataset 2 (Left- panel; Fig.~\ref{fig_vel2}). Fig.~\ref{fig_data2} shows the intensity, Doppler velocity, and FWHM maps for Mg\,{\sc ii} k (Fig.~\ref{fig_data2}; left-column), C\,{\sc ii} (Fig.~\ref{fig_data2}; middle column), and Si\,{\sc iv} (Fig.~\ref{fig_data2}; right-column). 
The Doppler velocity maps (middle-panel in each column; Fig.~\ref{fig_data2}) exhibits that one of the footpoint is red-shifted while the other is blue-shifted. 
The upflowing plasma (i.e., the blue-shifted end) is falling to the other end, which causes red-shifts there. The corresponding values of Doppler velocity are shown by the colorbars in each panel. 
The line-widths are represented by FWHM maps in the bottom panels of Fig.~\ref{fig_data2} for Mg\,{\sc ii} k3 (left-cloumn), Fig.~\ref{fig_data2} for C\,{\sc ii} (middle-column), and Fig.~\ref{fig_data2} for Si\,{\sc iv} (right-column) where the increased linewidth at the red-shifted footpoint might be due to downflowing plasma (Tian et al. 2008; Tian et al. 2009). 

\textbf{Variation of Doppler velocity:}
The locations used for Doppler velocity estimations are labelled as B1, B2, B3, B4, B5, B6 and R1, R2, R3 at the blue-shifted and red-shifted footpoints of the cool loop system for Dataset 2 shown in the left panel of of Fig.~\ref{fig_vel2}. The top-right panel shows the variation of Doppler velocity for Dataset 2 at the blue-shifted footpoints of the cool loop system for boxes shown in left panel of Fig.~\ref{fig_vel2}. The Doppler velocity of the Ni\,{\sc i} line is very low in all boxes showing almost no upflows or downflows (-0.07 {\rm to} 0.44) $\km\s^{-1}$ in the photospheric region. These flows show small red-shifts/blue-shifts for selected different boxes at the formation temperature of Mg\,{\sc ii}\,k. All these flows show considerable blue-shifts (-0.50 {\rm to} -2.74) $\km\s^{-1}$ at C\,{\sc ii}. The Doppler velocity at Si\,{\sc iv} shows high blueshifts for B4, B5, and B6 having maximum value of -13.83$\km\s^{-1}$. However, the boxes B1, B2, and B3 shows relatively small upflows having velocity (-4.13 {\rm to} -5.47) $\km\s^{-1}$. The corresponding error bars (1-$\sigma$ error) are overplotted, however, they are not visible due to their very small amplitude.

We have also considered three locations within the the redshifted footpoints of this cluster of cool loops, which are labelled  as R1, R2, and R3. The Doppler velocities increase for different spectral lines showing the increase in plasma downflows as we go
higher up in the solar atmosphere as shown in the bottom-right panel of Fig.~\ref{fig_vel2}. 
The Doppler velocity for Ni\,{\sc i} line shows small flows having range  (-0.08 {\rm to} 0.04) $\km\s^{-1}$. Mg\,{\sc ii}\,k line has relatively small downflows having Doppler velocity range of (0.65 {\rm to} 3.61) $\km\s^{-1}$. The downflows are then increased as we go higher to the formation height of C\,{\sc ii} having Doppler velocity range of (1.01 {\rm to} 7.84) $\km\s^{-1}$. These values reach to the maximum values of (5.11 {\rm to} 11.73) $\km\s^{-1}$ for Si\,{\sc iv} line. The maximum activity is found in the upper chromosphere/TR interface where the downflows become large and create excess widths at the associated footpoints.

\textbf{Line widths of Si\,{\sc iv} at red and blue-shifted footpoints of cool loops:}
We have also investigated the FWHM of the Si\,{\sc iv} line within all boxes (blue as well as red) for the
second data set. FWHM of the blue boxes of second observation is 0.107 \AA~(B1), 0.105 \AA~(B2), 0.108 \AA~(B3), 0.094 \AA~(B4), and 0.131 \AA~(B5), and 0.141 \AA~(B6). While, the red boxes have widths of 0.200 \AA~(R1), 0.186 \AA~R2), and 0.187 \AA~(R3) indicating higher line-widths at the red-shifted locations. A similar trend of FWHM in blue and red-boxes are also found in this observation as the dataset 1, therefore, the possible physical implications remain same.


\begin{figure*}
\mbox{
\hspace{-1.0cm}
\includegraphics[trim=6.0cm 1.0cm 5.0cm 0.5cm,scale=0.9]{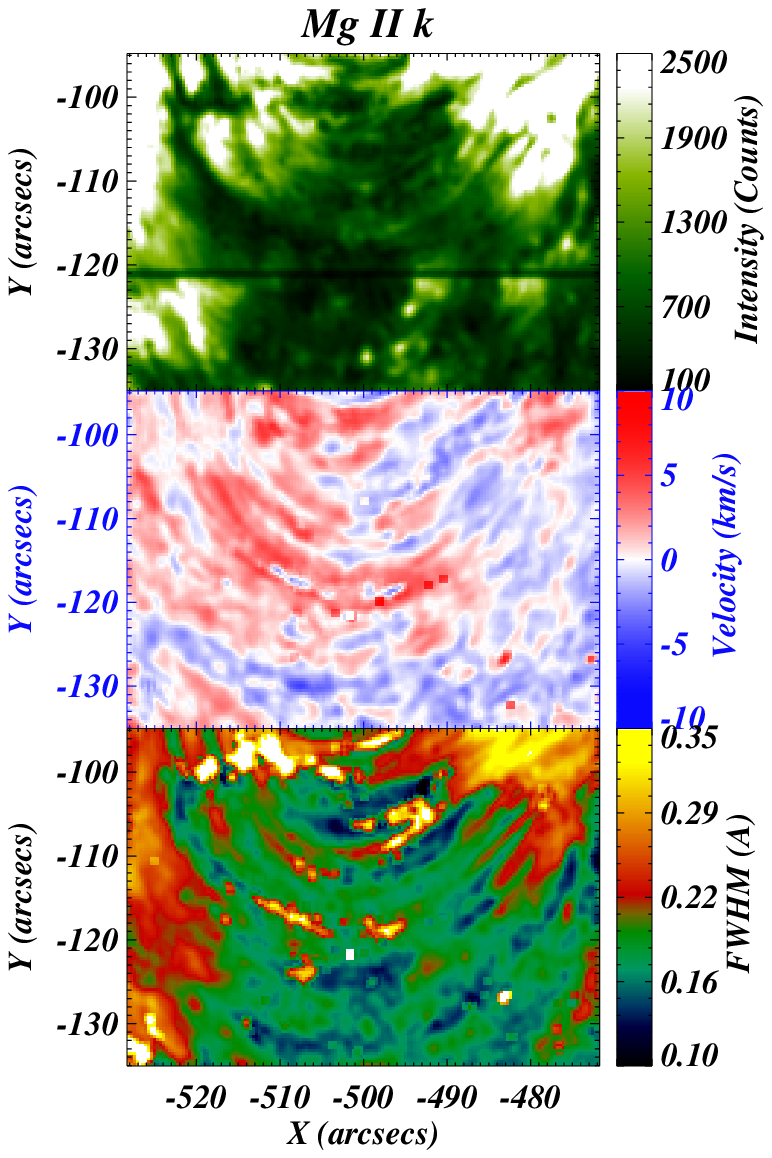}
\includegraphics[trim=5.0cm 1.0cm 5.0cm 0.5cm,scale=0.9]{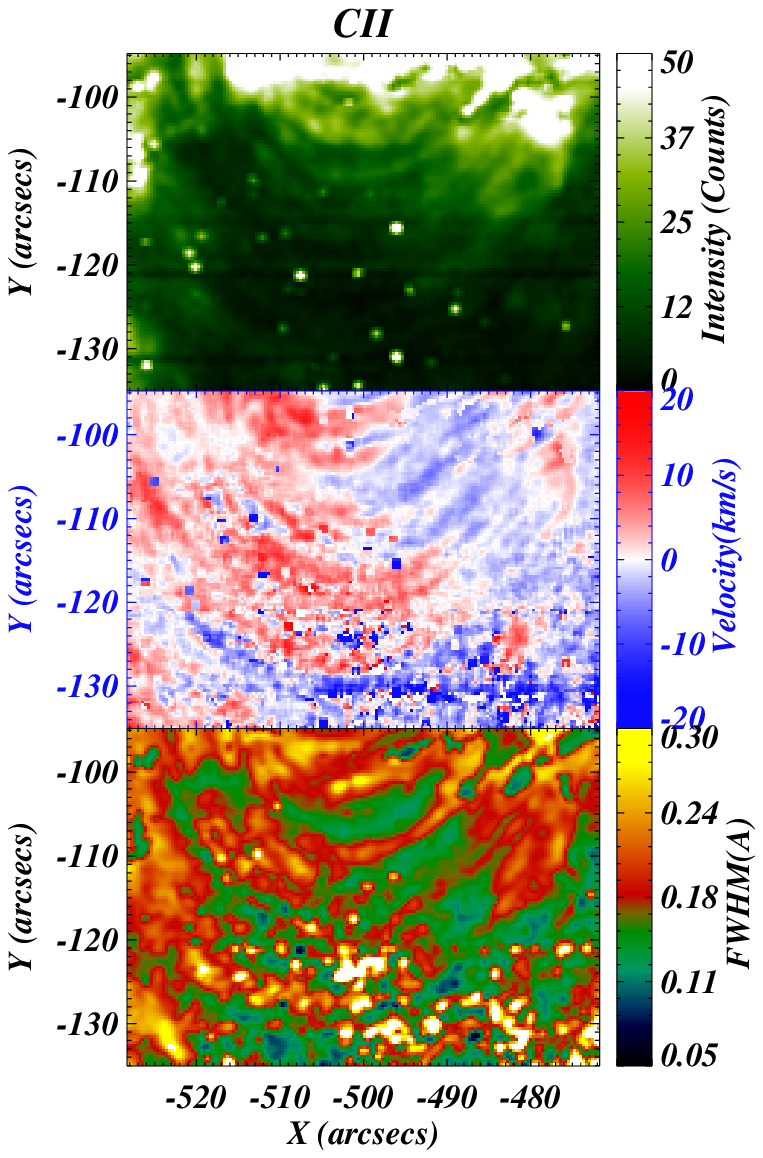}
\includegraphics[trim=5.0cm 1.0cm 5.0cm 0.5cm,scale=0.9]{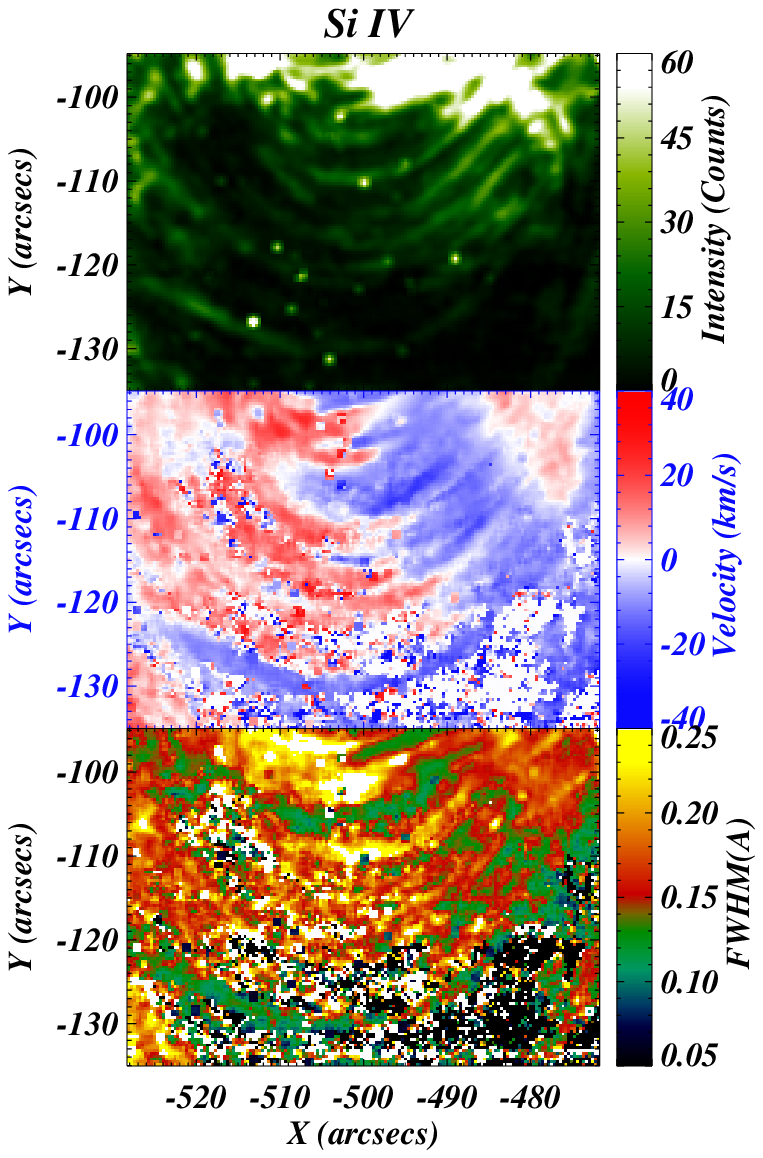}}
\caption{Intensity, Doppler velocity, and Full width at half maximum (FWHM) maps
of Mg\,{\sc II}\,k (2796.2~\AA) ,
C\,{\sc II} (1334.53~\AA), and
Si\,{\sc IV} (1402.77~\AA)
lines are shown for the Dataset 3
in the left, middle, and right columns respectively.}
\label{fig_data3}
\end{figure*}
\begin{figure*}[h]
\centering
\mbox{
\includegraphics[trim=2.5cm 5.0cm -8.0cm -0.5cm,scale=0.45]{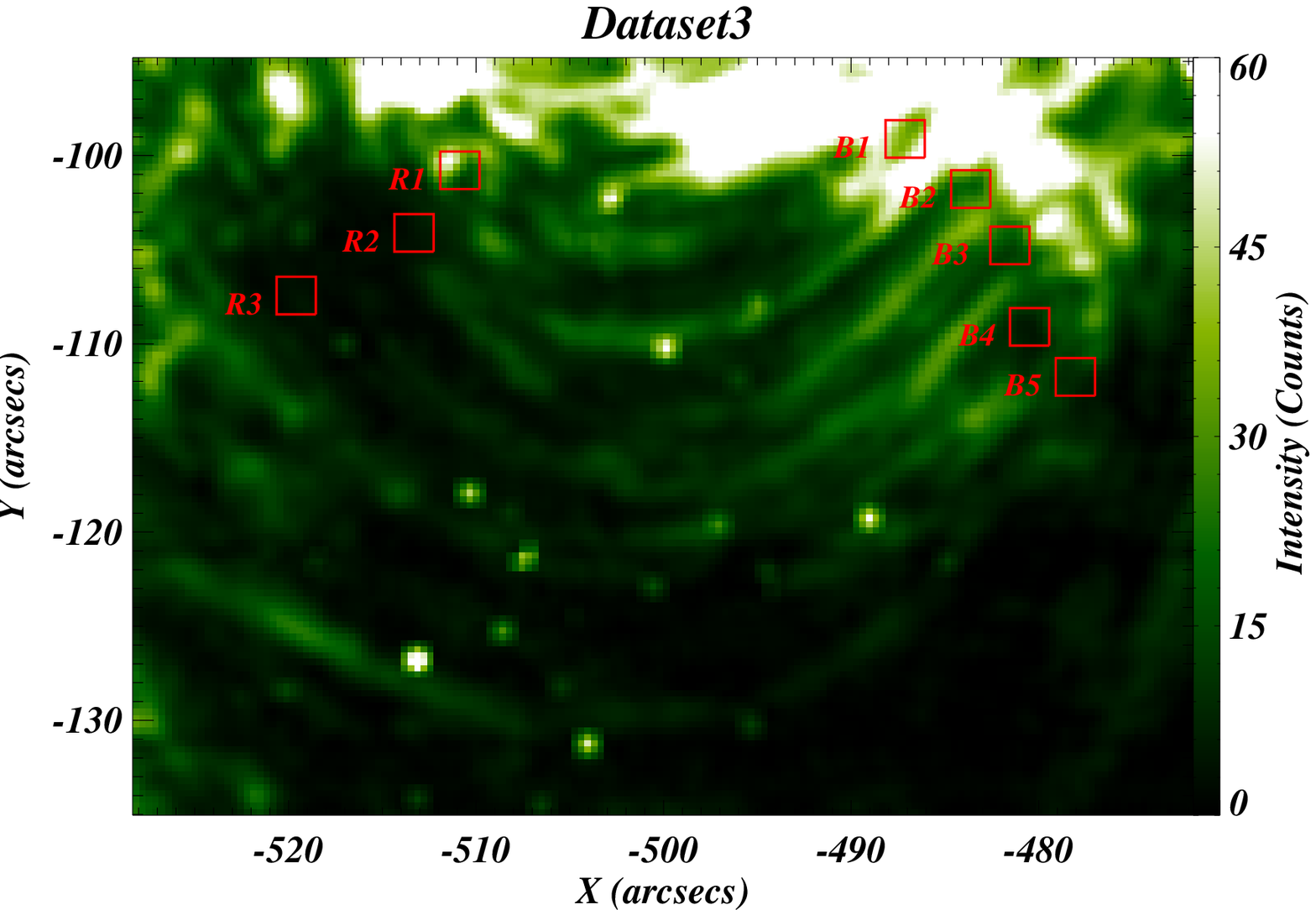}
\includegraphics[trim=8.0cm 0.5cm -1.0cm -1.0cm,scale=0.45]{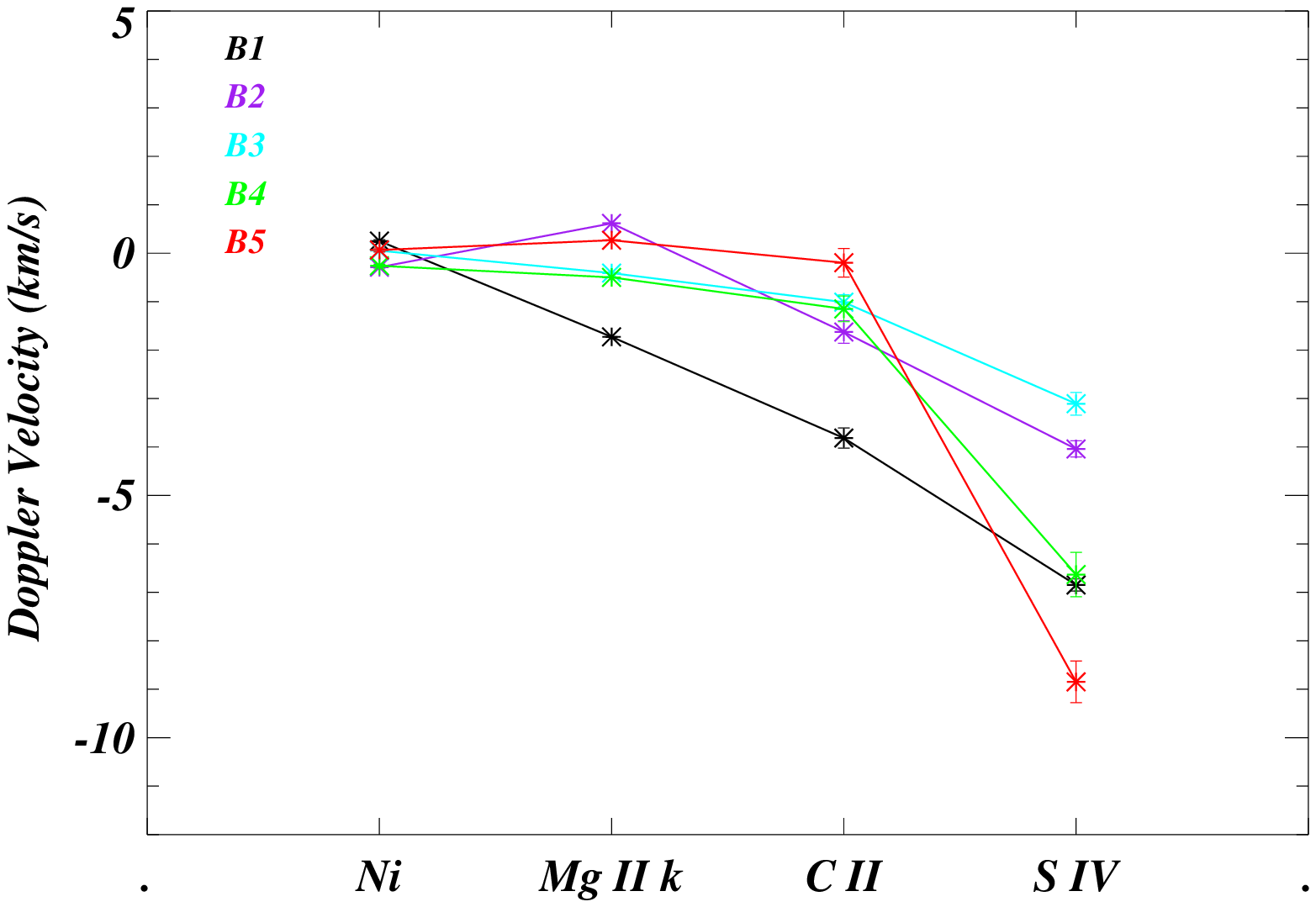} }
\mbox{   
\includegraphics[trim=-15.0cm 0.5cm -1.0cm 0.0cm,scale=0.45]{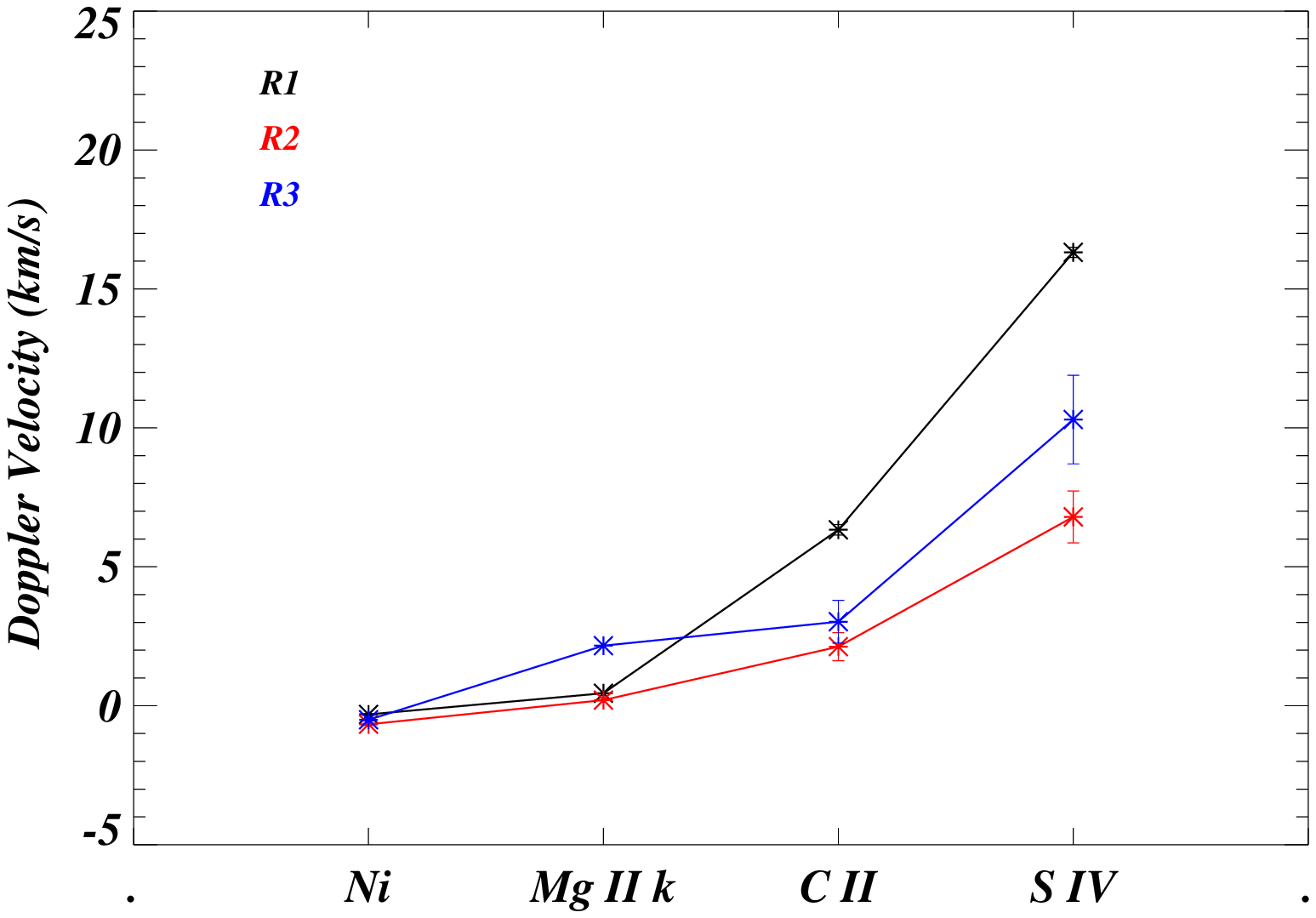}}     
\\
     
\caption{Left panel: The intensity map of Si\,{\sc IV} (1402.77~\AA) line with the boxes overlaid showing different locations at the footpoints of the cool loop systems. Top-right panel: The variation of Doppler velocity with the formation heights of different spectral lines for different boxes at the blueshifted footpoint indicated in the left panel. Bottom-right panel: The variation of Doppler velocity with the formation heights of different spectral lines for different boxes at the redshifted footpoint indicated in the left panel.}
\label{fig_vel3}

\end{figure*}

\subsection{Dataset 3}
\label{ss.3}
\textbf{Identification of cool loops:}
The right-column of Fig.~\ref{fig_mosaic1} also shows the mosaic of the intensity 
images for different spectral lines for Dataset 3 along with the magnetic polarities 
at the footpoints shown by LOS magnetogram (Fig.~\ref{fig_mosaic1}; panel I). The loop strands are visible as clearly resolved structures in Si\,{\sc IV} line intensity (Fig.~\ref{fig_mosaic1}; panel L) while Mg\,{\sc ii}\,k (Fig.~\ref{fig_mosaic1}; panel J) and C\,{\sc ii} (Fig.~\ref{fig_mosaic1}; panel K) shows chromospheric/TR signatures and the loop strands appear fuzzier. The loop structures are thus shown at different wavelengths corresponding to different formation heights in the solar atmosphere. The LOS magnetogram indicates the opposite magnetic polarities at the footpoints of the cool loop system.

\textbf{Parametric maps for the third dataset:}
Fig.~\ref{fig_spec3} represents average profiles of different spectral lines correspond to the box labelled as B3 for Dataset 3 (Left-panel; Fig.~\ref{fig_vel3}).  Fig.~\ref{fig_data3} shows the intensity, Doppler velocity, and FWHM maps for Mg\,{\sc ii} k (Fig.~\ref{fig_data3}; left-column), C\,{\sc ii} (Fig.~\ref{fig_data3}; middle-column), and Si\,{\sc iv} line (Fig.~\ref{fig_data3}; right-column). The plasma emissions in the intensity maps (Top-panels; Fig.~\ref{fig_data3}) are similar to the direction of plasma flows in the Doppler velocity maps (Middle-panels; Fig.~\ref{fig_data3}) having blue-shifts at one end and red-shifts at the other end. The linewidths in the bottom panels of Fig.~\ref{fig_data3} show the mass motions signatures of the plasma having high values of FWHM at the red-shifted footpoint indicating the heating caused by downfalling plasma (Tian et al. 2008; Tian et al. 2009).

\textbf{Variation of Doppler velocity:}
In Fig.~\ref{fig_vel3}, the top-right panel shows the variation of Doppler velocity for Dataset 3 at the blue-shifted footpoints of the cool loop system for boxes shown in Fig.~\ref{fig_data3}. The Doppler velocity of the Ni\,{\sc i} line is very low in all boxes showing almost no upflows or downflows (-0.05 {\rm to} -0.28) $\km\s^{-1}$ in the photospheric region. These flows show small red-shifts and blue-shifts for different boxes at the formation temperature of Mg\,{\sc ii}\,k lying in the range (-1.72 {\rm to} 0.61) $\km\s^{-1}$. Interestingly, all these boxes show considerable blue-shifts (-0.19 {\rm to} -3.81) $\km\s^{-1}$ at C\,{\sc ii}. The Doppler velocity variation at Si\,{\sc iv} shows high blueshifts for having maximum value of -8.84$\km\s^{-1}$ for B5. 

In the similar fashion as we did in the previous cases, we have also considered three boxes in the redshifted region near other footpoints of this cluster of cool loops. The redshifts increase for different spectral lines showing the increase in plasma downflows as we go higher up in the solar atmosphere as shown in the bottom-right panel of Fig.~\ref{fig_vel3}. The maximum activity is found in the upper chromosphere/TR interface where the downflows become large and create excess widths at the associated footpoints. The variation of Doppler velocities above redshifted footpoint for different boxes (R1, R2, and R3) shows the increasing trend as observed in previous datasets for downflowing velocities at different formation heights of the spectral lines where the maximum values reaches the formation height of Si\,{\sc iv} having Doppler velocity range (6.79 {\rm to} 16.31) $\km\s^{-1}$.

\textbf{Line widths of Si\,{\sc iv} at red and blue-shifted footpoints of cool loops:}
We have also investigated the FWHM of Si\,{\sc iv} line within all boxes (blue as well as red) for the
third data set. The FWHM of the blue boxes of third observation is 0.149 \AA~(B1), 0.135 \AA~(B2), 0.134 \AA~(B3), 0.125\AA~(B4), and 0.126 \AA~(B5). While, the red boxes have the widths of 0.182 \AA~(R1), 0.139 \AA~R2), and 0.172 \AA~(R3). Possible physical explanations for such excess width have already been mentioned for the datasets 1 and 2.


\section{Discussions and Conclusions} 
\label{s.concl}
We have studied various cool loop systems in order to understand the
co-spatial variations of
Doppler velocities at various locations with blue-shifted footpoints as well as 
red-shifted footpoints for different spectral lines having different formation heights. 
The Doppler variation shows the increasing height trend in the solar atmosphere.

The photospheric, as well as chromospheric regions, show very small
blueshifts of $\sim (+1~{\rm to}~-1) \km\s^{-1}$,
which exhibits almost no upflow signatures till Mg\,{\sc II}~k.
On the contrary, the C\,{\sc ii} lines are significantly
blue-shifted by $\sim (-1~{\rm to}~-6) \km\s^{-1}$.
The blue-shifted nature becomes more prominent at the formation temperature
of Si\,{\sc IV}.
In the Dataset 1 and Dataset 2, upflows are 
indicated above C\,{\sc II} formation height at the blue-shifted footpoints which is well supported 
by the downflowing plasma at the red-shifted footpoints. However, for the Dataset 3, these 
upflows start below C\,{\sc II} formation height (upper chromosphere) and then driven by plasma inertia 
to higher heights till Si\,{\sc IV} and shows small downflows at C\,{\sc II} at red-shifted footpoints.
All the blue-shifted locations for the same Dataset follow similar pattern but have different range of Doppler velocities
showing the localised impulsive events. These flows initiated by short impulsive events which are highly localised in the 
upper chromosphere or lower TR.
Such localised impulsive events above or below the C\,{\sc II} formation height 
make the loops thermally unstable.

So, the significant plasma upflows take place between the formation heights of C\,{\sc II} and Si\,{\sc IV} lines
for Dataset 1 and Dataset~2 while it takes place below the formation heights of C\,{\sc II}.
The transition of small up-flow to high up-flow velocities takes place near
the formation height of C\,{\sc II} line,
which justifies the origin site of plasma flow in these cool loops.

Different plausible mechanisms have been already discussed for the origin of flows in coronal loops,
e.g., siphon flow (Hood \& Priest 1981),
downward propagating acoustic waves (Hansteen 1993),
explosion below corona (Teriaca et al. 1999),
acoustic waves (Taroyan et al. 2005)
and nanoflare driven chromospheric evaporation (Patsourakos \& Klimchuk 2006).
However, the occurence of implusive heating involving magnetic reconnection has recently been
reported in support of the plasma flows
in the cool loop systems (Huang et al. 2015).

Warren et al. (2002) have suggested the impulsive heating of active region hot loops
using TRACE observations. 
Spadaro et al. (2003) have also explained the flows in the coronal loops 
due to transient heating near the chromsophere.
The outflow locations at the footpoints in cool loop systems 
have higher emissions indicating the dynamical nature of the loops.
Localised impulsive energy release might be driving such flows since
there are no flows at a few footpoints of the cool loop system
(B1, B2, B3; Fig.~\ref{fig_vel2}) while other footpoints near
it shows upflows.
The heating due to impulsive energy release gets
deposited at that footpoint to drive the upflows.
That portion of the flowing plasma arriving at the opposite footpoint,
causes an enhancement of the linewidth.
Thus, the speculation of such loop systems driven by heating
pulses due to magnetic reconnection is supported by our observations.
The non-thermal broadening of Si\,{\sc iv} line at the 
blue-shifted footpoints of the cool-loop system is most likely associated 
with the ongoing transient energy release  and thereafter filling-up of the plasma 
in loop threads (e.g., Patsourakos \& Klimchuk 2006;
Hansteen et al. 2014; Huang et al. 2015; Polito et al. 2015). 
Red-shifted footpoints in the Si\,{\sc iv} line indicates that the TR plasma is flowing 
down on the other end of the loop system, and the excess non-thermal width may be associated 
with the variety of physical processes there, e.g.,  downfalling of the plasma (Tian et al. 2009) , 
nano-flare 
generated acoustic waves (Hansteen 1993), 
downward propagating pressure disturbances (Zacharias et al. 2018), etc. 
Exact answer is still an open question that what physical process(es) causes these 
TR footpoints of the cool loop threads with excess line-width.

Since the transition of plasma upflows start at the TR
(near the C\,{\sc ii} line formation height), they can transport
mass and energy directly to the inner corona above the cool loop systems. Therefore, such loop
systems play a significant role in the mass, energy and momentum transport.
They may be the lower counterparts of high temperature
coronal loops. High speed plasma flows in the magnetic domain
of the loops have been observed in the solar atmosphere
(Harra et al. 2008). However, there are several unclear
proposition on the origin of such fast flows.
The possible transport of the plasma from the low-lying cool loop systems
(the one we observed) to the higher magnetic field configuration
where the plasma acceleration starts, may be the most likely source.
Such plasma flows, after their acceleration in the higher atmosphere,
may provide the plasma for the
solar wind, if the open field lines exist in the vicinity of the
footpoints of such loops,
though, such case has not been observed in our case.

In conclusion, our work emphasizes the significance of plasma flows at different formation heights of the spectral lines corresponding to different regions of the solar atmosphere in
a cool loop system. The height-dependent trend of the Doppler velocities explores the region where these flows start to show prominent upflows. Such flows do not start from the photosphere.
On the contrary, they start from the upper chromosphere and TR where 
transition from small blueshifts to significantly high blueshifts have been
identified in such a system. Future investigations would be required
for comparing the Doppler velocity patterns in cool and hot coronal
loops using multi-line spectroscopy in order to differentiate the underlying
causes and different origins of the plasma flows there.

\section{Acknowledgements}
We thank referee for his/her constructive comments which improved the manuscript.
One of us (Yamini K. Rao) is fully supported by the financial grant from the ISRO RESPOND project. AKS acknowledges UKIERI project grant. PK acknowledge the grant Rozvoj JU -- Mezinarodni mobility--CZ.02.2.69/0.0/16$\_$027/0008364. We acknowledge the use of IRIS observations. IRIS is a NASA small explorer mission developed and operated by Lockheed Martin Solar and Astrophysics Laboratory (LMSAL) with mission operations executed at NASA Ames Research Center and major contributions to downlink communications funded by the Norwegian Space Center (NSC, Norway) through an ESA PRODEX contract.


\appendix 
\renewcommand\thefigure{\thesection.\arabic{figure}}
\section{Fitting Note on Optically thick lines}\label{section:append}
Mg\,{\sc ii} k 2796.35~{\AA} and C\,{\sc ii} 1334.53~{\AA} are optically thick spectral lines, which can have single or double peaks or even more (cf; Fig~\ref{fig:append_a2}) (Leenaarts et al. 2013; Rathore et al. 2015). In the present observations, we have found that a major fraction of of Mg\,{\sc ii} k 2796.35~{\AA} has double peak profiles (i.e., k2v and k2r) with associated minimum between these two peaks (i.e., k3). However, in case of sunspot umbra the Mg\,{\sc ii} k are single peak profiles as reported by Tian et al. (2014). We have used two Gaussian (one positive and another negative) along the straight line to fit the Mg\,{\sc ii} k 2796.35~{\AA} line. Our fitting model is similar to the one used by Schmit et al. (2015). We have used straight line to fit the continuum while Schmit et al. (2015) have used a polynomial.  
\setcounter{figure}{0}   

\begin{figure*}
\vspace{-1.2cm}
\mbox{
\includegraphics[trim=1.5cm 3.5cm 2.0cm 3.0cm,scale=1.1]{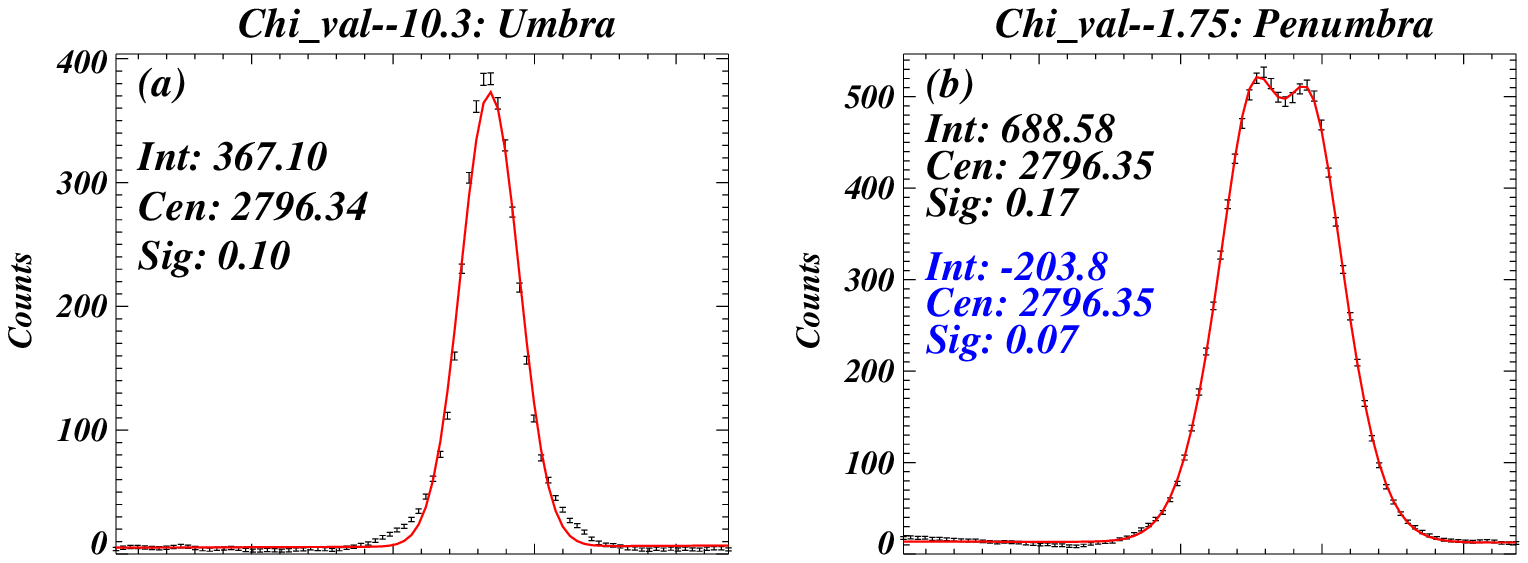}
}
\mbox{
\includegraphics[trim=1.5cm 3.5cm 2.0cm 3.0cm,scale=1.1]{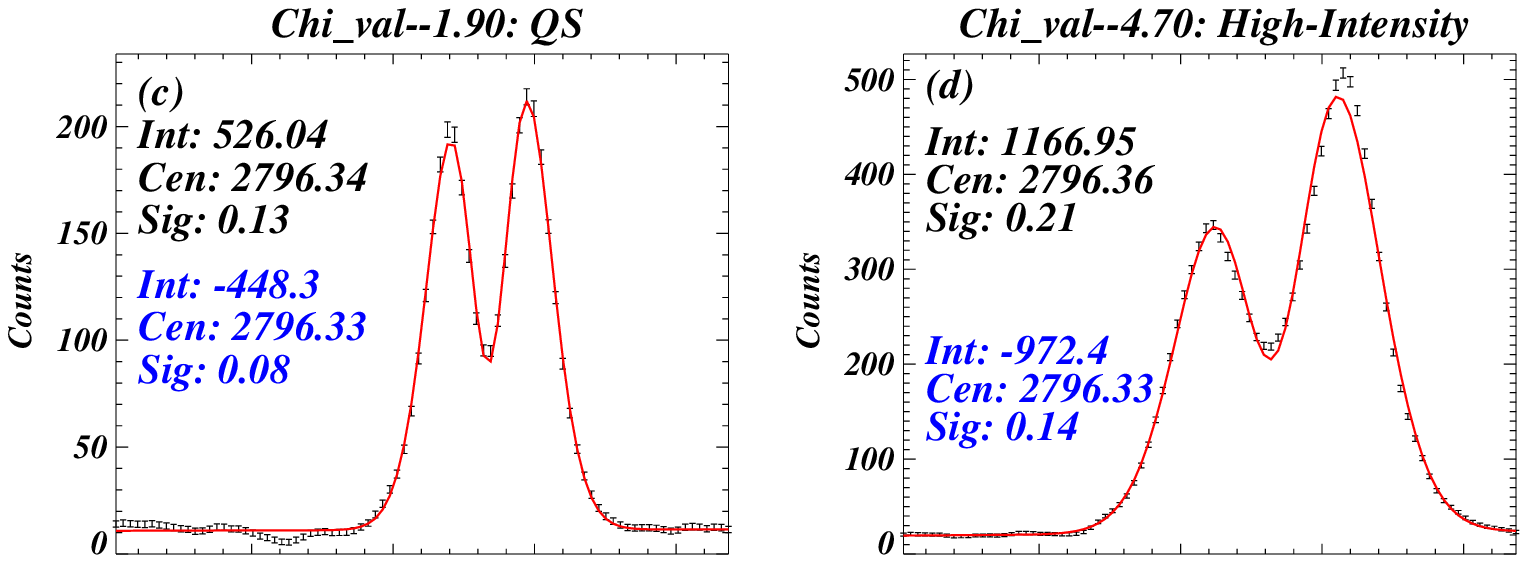}
}
\mbox{
\includegraphics[trim=1.5cm 3.0cm 2.0cm 3.0cm,scale=1.1]{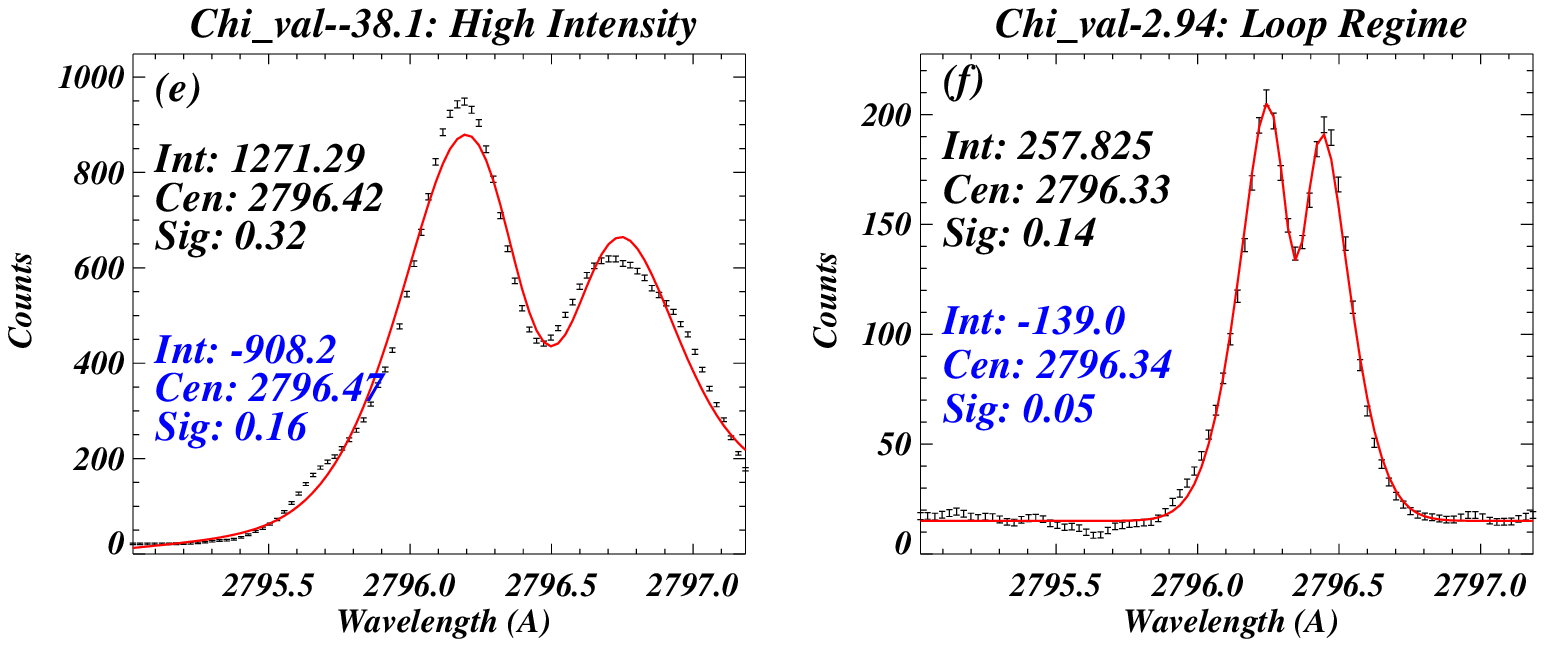}
}
\caption{Few samples of our fitting model on the observed profiles of Mg\,{\sc ii} k 2796.35~{\AA} within the different features, e.g., umbra (panel), penumbra (panel b), QS (panel c), high-intensity (panel d $\&$ e) and loop regime (panel f). The black shows the observed profiles while the red line is fitted model on the observed profiles. We have also displayed the parameters (peak intensity, centroid and Gaussian sigma) from positive Gaussian (black) and negative Gaussian (blue) in each panel. These samples illustrates the variations of Mg\,{\sc ii} k line within the different features in the observed region. The fitting model capture the line behaviour very well, which justify our fitting model for Mg\,{\sc ii} k line.}
\label{fig:append_a1}
\end{figure*}
\begin{figure*}
\mbox{
\includegraphics[trim=1.5cm 2.0cm 2.0cm 3.0cm,scale=1.1]{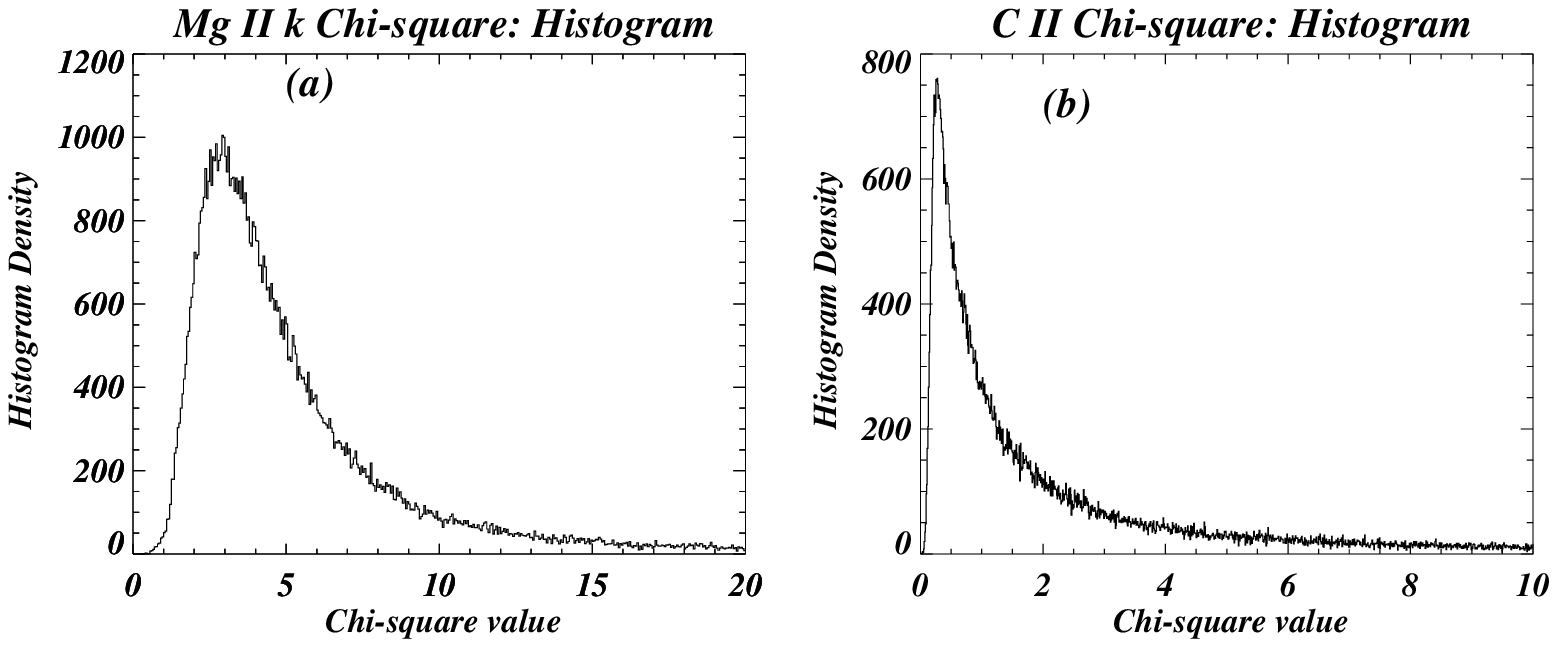}
}
\mbox{
\includegraphics[trim=1.5cm 2.0cm 2.0cm 3.0cm,scale=1.1]{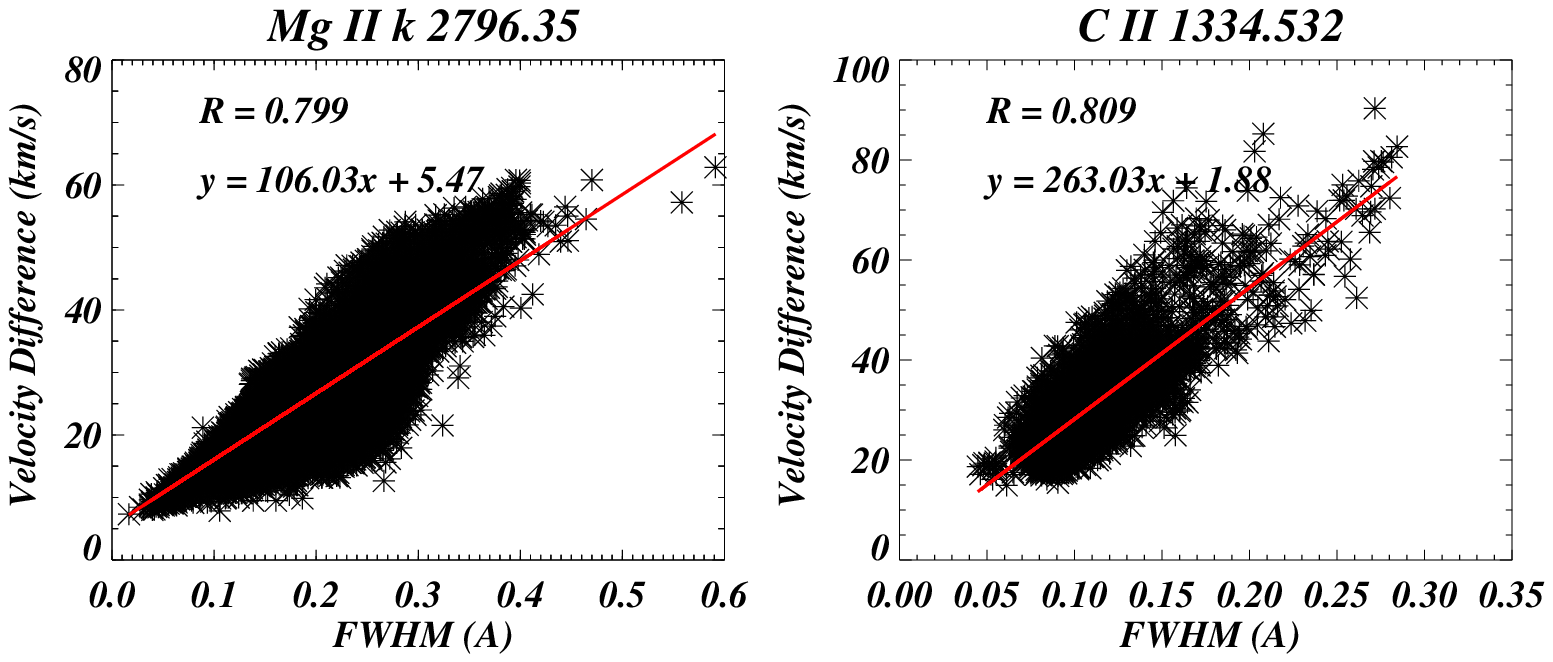}
}
\caption{Panel (a) shows the distribution of chi-square values for Mg\,{\sc ii} k, which are estimated for each fitted profile within the Dataset1. Statistical distribution peaks around 3.0 with its major extension up to 12, which shows that majority of profiles are fitted very well with the current model. Similarly, panel (b) shows the similar chi-square distribution for C\,{\sc ii}. It peaks around 1 with its major extension upto 5 justifies the fitting of C\,{\sc ii} line also. In bottom-left panel, we have shown the correlation between velocity difference (i.e., k2r-k2v) and FWHM of negative Gaussian. We have found very tight correlation between velocity difference and FWHM. Similarly, the bottom-right panel shows the same correlation but for C\,{\sc ii}, which is also very tight.}
\label{fig:append_a4}
\end{figure*}

Fig.~\ref{fig:append_a1} shows some sample fittings from various features within the observed region (i.e., Dataset1). Typical single peak profiles exist in the sunspot umbra (panel a) while a slight dip is visible within the penumbra atmosphere (panel b). We have shown the chi-squared value and the fitting parameters (positive Gaussian (black) and negative Gaussian (blue)) in each plot. However, in other features (i.e., QS, high-intensity regions (plages), and loops) the Mg\,{\sc ii} k profiles are the typical double peak profiles (panel c,d,e and f) as also reported in previous IRIS related works (Carlsson et al. 2015; Schmit et al. 2015; Kayshap et al. 2018). In case of panel (e), the profile comes from very high intensity/activity regions, which has high chi-square value. However, the observed profiles are well characterized by the fitting model. In rest of the cases, the chi-square values are in good agreement. Some very complex spectral profiles also exist, which are shown in Fig~\ref{fig:append_a2}.
\begin{figure*}
\mbox{
\includegraphics[trim=1.5cm 3.5cm 2.0cm 3.0cm,scale=1.1]{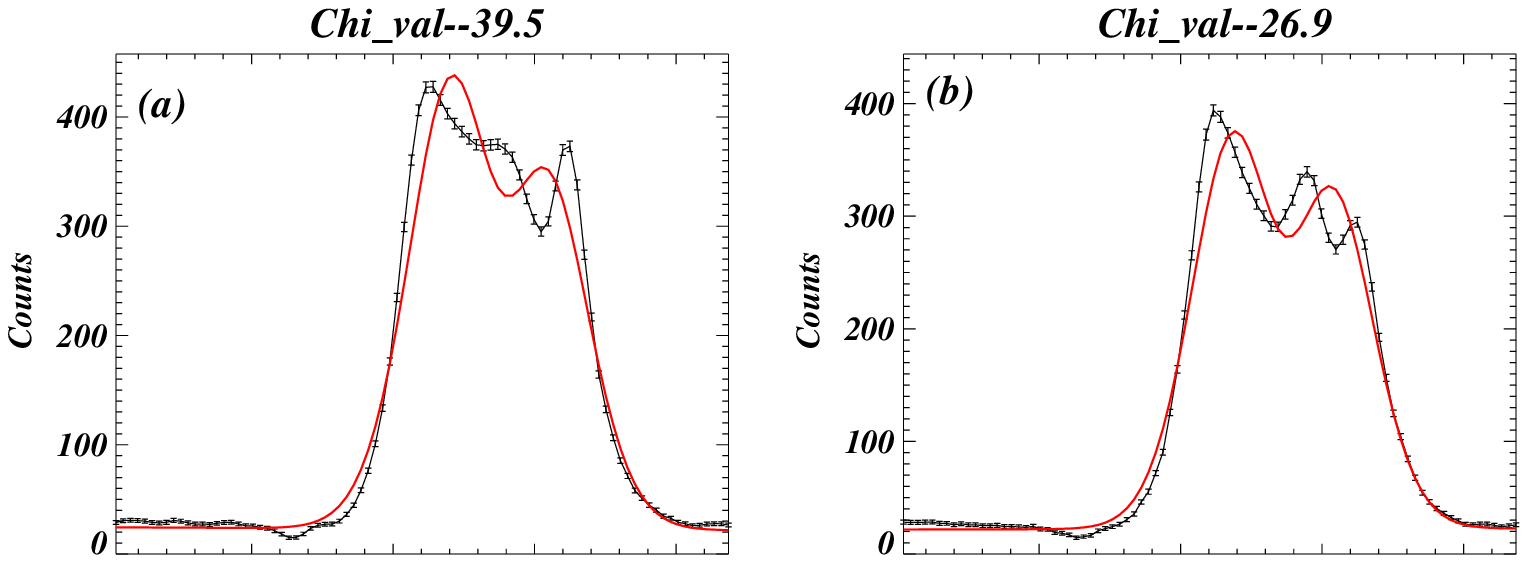}
}
\mbox{
\includegraphics[trim=1.5cm 3.0cm 2.0cm 3.0cm,scale=1.1]{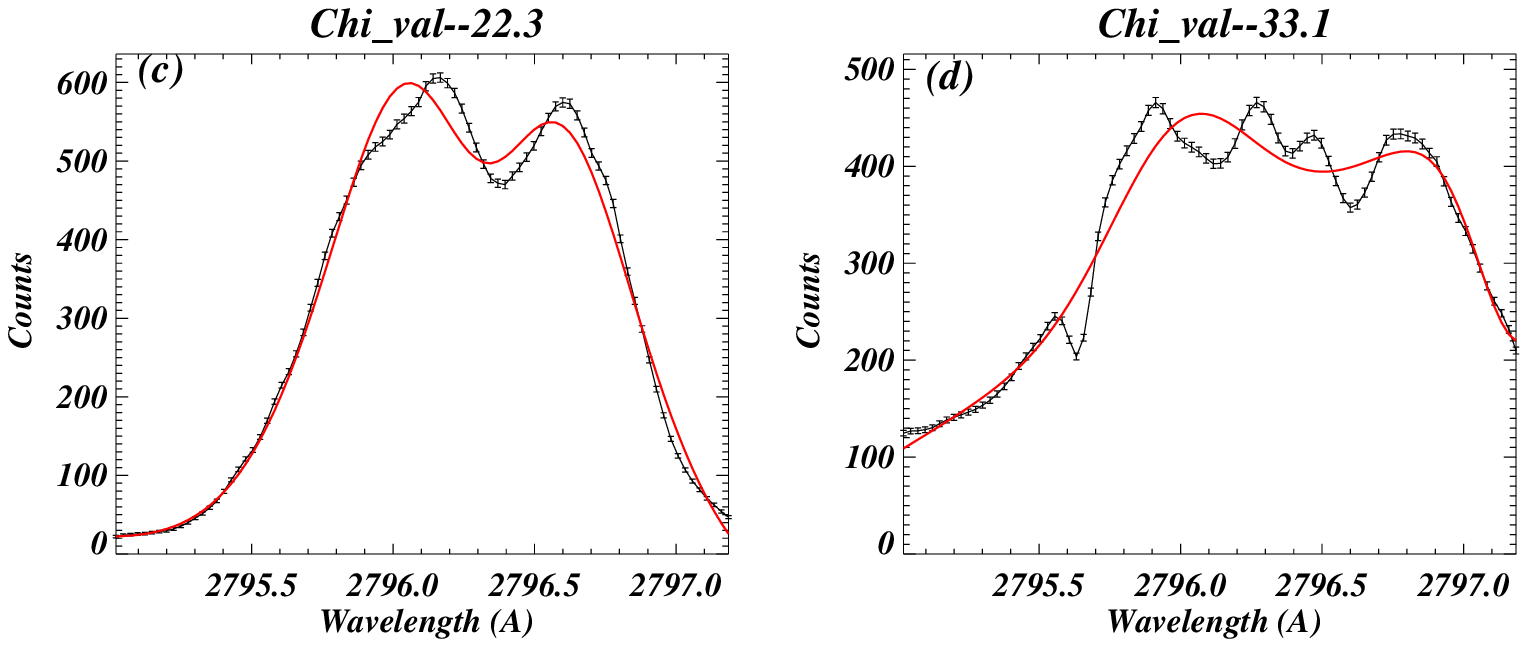}
}
\caption{In this figure, we have shown some very complex Mg\,{\sc ii} k profiles occuring in the Dataset1. These complex profiles have 3, 4 or 5 peaks, which can not be fitted with the current fitting model. However, these profiles are very rare and we did not taken these profiles in our further analysis.}
\label{fig:append_a2}
\end{figure*}
These complex spectral profiles have 3 or more peaks, which cannot be characterized with the present model (see; fitted red lines on the observed profiles; Fig~\ref{fig:append_a2}). However, it should be noted that such profiles are very rare in the present observations. Therefore, we have not considered such profiles in our analysis. We have estimated the chi-squared value for each spectral profile in all the three observations. The statistical distribution of chi-square values of Mg\,{\sc ii} k 2796.35~{\AA} is shown in the panel (a) of Fig.~\ref{fig:append_a4}. It is clearly visible that the chi-sqaure value is less than 10.0 for most of spectral profiles, which justify our fitting model for Mg\,{\sc ii} k 2796.35~{\AA}. Similar type of chi-square distributions are also found for other datasets (Dataset 2 and Dataset 3), which are not shown here. \\
\begin{figure*}
\mbox{
\includegraphics[trim=1.5cm 3.5cm 2.0cm 3.0cm,scale=1.1]{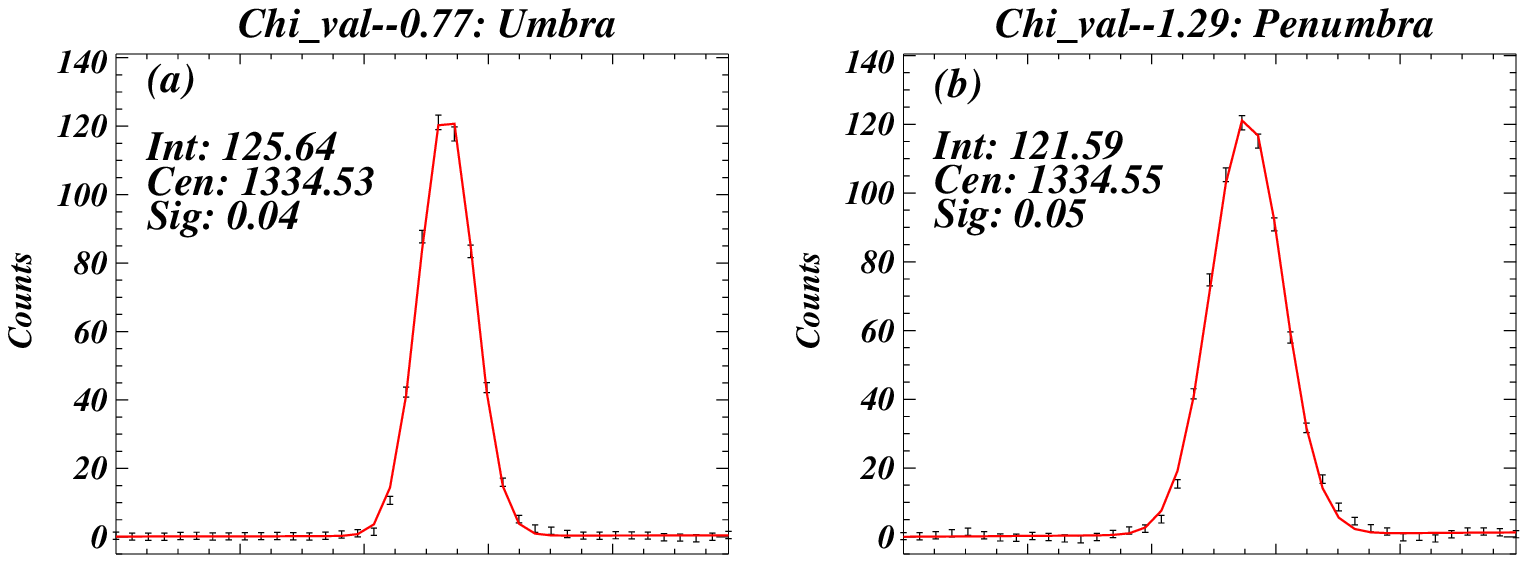}
}
\mbox{
\includegraphics[trim=1.5cm 3.5cm 2.0cm 3.0cm,scale=1.1]{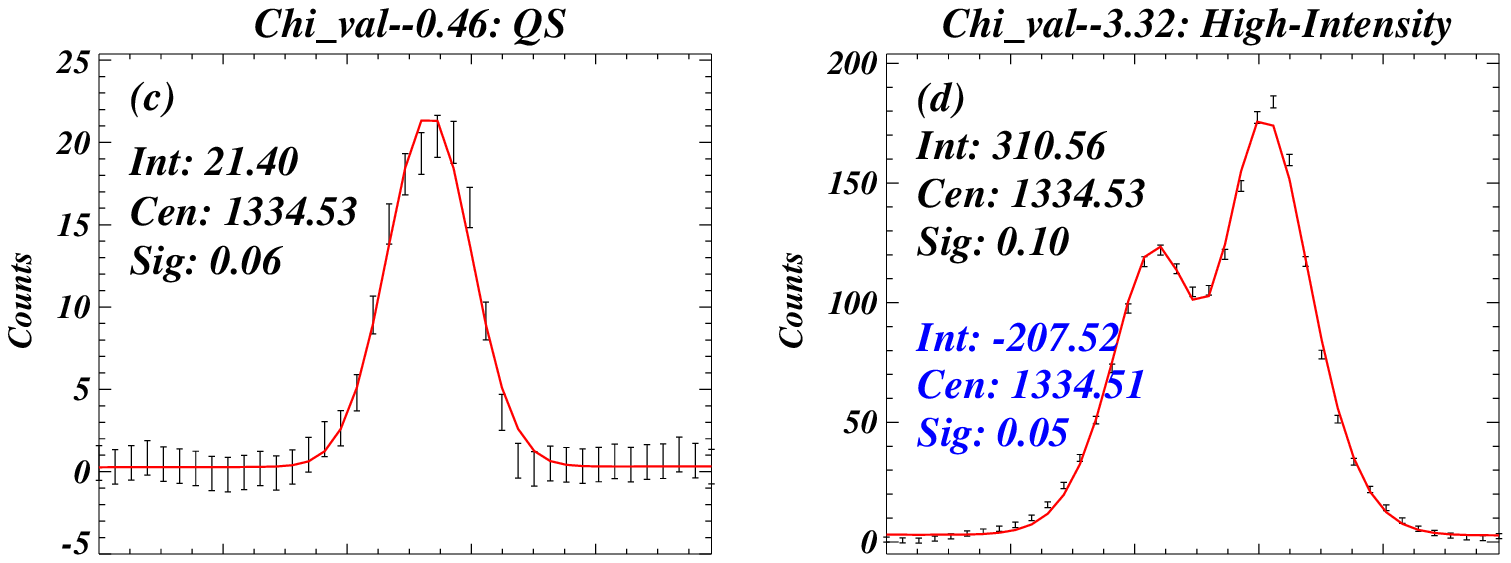}
}
\mbox{
\includegraphics[trim=1.5cm 2.5cm 2.0cm 3.0cm,scale=1.1]{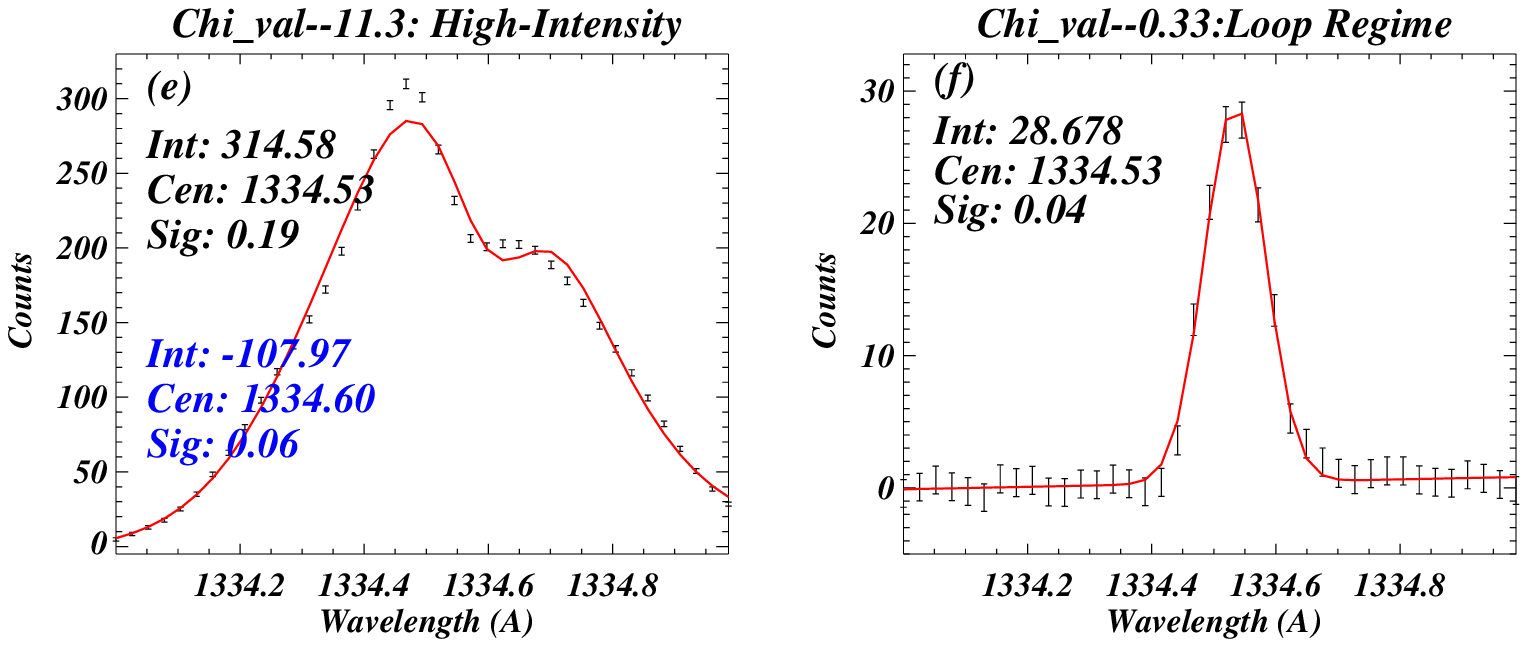}
}
\caption{Sample profiles along with their fitting as we shown in Fig~\ref{fig:append_a1} but for C\,{\sc ii} 1334.53~{\AA}. Interestingly, C\,{\sc ii} 1334.53~{\AA} is single peak profiles in all features expect high-intensity area of the observations. The low value of chi-square justifies the reliability of our fitting model for C\,{\sc ii} 1334.53~{\AA}}
\label{fig:append_a3}
\end{figure*}
Two resonance C\,{\sc ii} lines (C\,{\sc ii} 1334.53~{\AA} and C\,{\sc ii} 1335.71~{\AA}) are present in the IRIS FUV filter observation. In general, C\,{\sc ii} 1335.71~{\AA} is stronger than C\,{\sc ii} 1334.53~{\AA}, however, the stronger line is blended with another nearly present C\,{\sc ii} 1335.66~{\AA} line. Therefore, we have used weak but clean C\,{\sc ii} 1334.53~{\AA} in the present analysis. Considering C\,{\sc ii} lines  as already discussed, we have used the same fitting model as for Mg\,{\sc ii} k 2796.35~{\AA} double peak profiles. The single peak profiles are fitted by single Gaussian. We have used the averaged spectral profile (with $\pm$1 at every pixel location) as the exposure time is very low in these observations, so the counts are low in this line. We have some sample spectral fitting for Mg\,{\sc ii} k 2796.35~{\AA} and C\,{\sc ii} 1334.532~{\AA} lines. However, it should be noted that C\,{\sc ii} 1334.53~{\AA} spectral line (weak but clean) has more single peaks (i.e., panel a-umbra, panel b-penumbra, panel c-QS and panel f-loop-regime) compared to C\,{\sc ii} 1335.77~{\AA} line (Rathore et al. 2015). Fig.~\ref{fig:append_a3} shows some spectral profiles from various features (as we shown in case of Mg\,{\sc ii} k; Fig.~\ref{fig:append_a1}) from the Dataset1 observation. However, the double peak nature of C\,{\sc ii} 1334.53~{\AA} line becomes prominent in the high-intensity regions (panel d and e). We have found that almost 8 \% profiles are double peaks profiles, which are present mostly in the high intensity areas. The chi-square distribution of C\,{\sc ii} 1334.53~{\AA} fitting, which is shown in the panel (b) of Fig.~\ref{fig:append_a4}. We find that the chi-square value is less than 2.0 for majority of the profiles. The similar type of chi-square distributions is found for other datasets (i.e., data-set 2 and data-set 3), which are not shown here. \\
We have used negative Gaussian for the estimation of Doppler velocity and FWHM for these optically thick lines. In this fitting model, the negative Gaussian basically captures the dip region in between two peaks. The dip-region is formed higher up in the chromosphere (Leenaarts et al. 2013; Rathore et al. 2015). As already noted, these optically thick lines can also have single peak profiles (specifically, C\,{\sc ii} 1334.532~{\AA} line), Therefore, we do not have the negative Gaussian parameters there. So, the parameters from single Gaussian (as profiles is single peak) are used at those locations to create the corresponding intensity, Doppler velocity and FWHM maps (cf., Figs.\ref{fig_data1}, \ref{fig_data2} and \ref{fig_data3}). 
Since negative Gaussian captures the variations of minimum region (dip) between two peaks, the spectral distance between k2v and k2r also varies as per the nature of spectral profiles. The negative Gaussian captures the activity of dip regions, therefore, this spectral distance (k2r and k2v) should reflect in the width of (i.e., FWHM) of negative Gaussian. Therefore, we should expect the positive correlation between the spectral distance between k2r and k2v (or the Doppler velocity difference) FWHM of negative Gaussian. Schmit et al. (2015) have estimated different widths for Mg\,{\sc ii} h 2803.53~{\AA}, i.e., distance between h1v $\&$ h1r and distance between h2v $\&$ h2r. They have fitted positive as well as negative Gaussian on the observed profiles using the fitting model (fitted profile) to draw the various peaks and dips (i.e., h1v, h1r, h2v,h2r and h3) from the fitting model. They have then estimated these two types of width (i.e., h1r-h1v and h2r-h2v). To validate our analysis, we have also used the fitting profiles to draw the k2v and kr peaks from the double peaked profiles for Mg\,{\sc ii} k and C\,{\sc ii} 1334.53~{\AA}. Finally, we have checked the correlation between Doppler velocity difference (k2r-k2v) and the FWHM of negative Gaussian for Mg\,{\sc ii} k 2976.35~{\AA} as well as C\,{\sc ii} 1334.53~{\AA}. The estimated correlations are shown in the Fig.~\ref{fig:append_a4}. Bottom-left panel shows the correlation for Mg\,{\sc ii} k 2796.35~{\AA} (panel (c); Fig~\ref{fig:append_a4}) while bottom-right panel shows this correlation for C\,{\sc ii} 1334.53~{\AA}. In case of Dataset 1, we have found a very well positive correlation for both lines (i.e., Mg\,{\sc ii} k) with the Pearson cofficient of 0.799. Similarly, the C\,{\sc ii} k line also shows very well correlation between velocity difference and FWHM of negative Gaussian with the Pearson coefficient of 0.809. We have also found the similar results for other datasets (i.e., Dataset 2 and Dataset 3), which are not shown here. So, this analysis validates that negative Gaussian basically captures the dynamics of the dip region (for Mg\,{\sc ii} k and C\,{\sc ii} lines), which originates from the upper chromopshere (Leenaarts et al. 2013; Rathore et al. 2015). Our adopted methodology is different from Schmit et al. (2015) to tackle the optically thick lines. Schmit et al. (2015) have utilized the best fitted profiles to estimate various parameters of Mg\,{\sc ii} h 2803.53~{\AA}, however, we have used the parameters from negative Gaussian to characterize the upper chromosphere. The motive behind the correlation (correlation between velocity difference and FWHM of negative Gaussian) is to emphasize that dynamics/activities are really captured by negative Gaussian. In addition, our adopted method also shows consistency with Schmit et al. (2015). 
\end{document}